\documentclass[final,5p,times,twocolumn]{elsarticle}

\usepackage{graphicx}
\usepackage{epsfig}
\usepackage{amssymb}
\usepackage{amsthm}
\usepackage{color}
\usepackage{multirow}
\usepackage{twoopt}
\usepackage{graphicx}
\usepackage[hyphens]{url}
\usepackage{amsmath}

\usepackage[pdftex,         %
            breaklinks=true,%
            colorlinks=true,%
            pdfauthor={Maurin},%
            pdftitle={USINE v3.4 and v3.5 for CPC}%
           ]{hyperref}

\usepackage[numbers]{natbib}
\biboptions{sort,numbers}
\newcommand{\ascii}{{\tt ASCII}}
\newcommand{\rootcern}{{\sc root cern}}
\newcommand{\usine}{{\sc usine}}

\def\pbar{$\overline{p}$}

\makeatletter
\newcommandtwoopt{\citeads}[3][][]{\href{http://adsabs.harvard.edu/abs/#3}{\def\hyper@linkstart##1##2{}\let\hyper@linkend\@empty\citealp[#1][#2]{#3}}}
\newcommandtwoopt{\citepads}[3][][]{\href{http://adsabs.harvard.edu/abs/#3}{\def\hyper@linkstart##1##2{}\let\hyper@linkend\@empty\citep[#1][#2]{#3}}}
\newcommandtwoopt{\citetads}[3][][]{\href{http://adsabs.harvard.edu/abs/#3}{\def\hyper@linkstart##1##2{}\let\hyper@linkend\@empty\citet[#1][#2]{#3}}}
\newcommandtwoopt{\citealpads}[3][][]{\href{http://adsabs.harvard.edu/abs/#3}{\def\hyper@linkstart##1##2{}\let\hyper@linkend\@empty\citealp[#1][#2]{#3}}}
\newcommandtwoopt{\citealtads}[3][][]{\href{http://adsabs.harvard.edu/abs/#3}{\def\hyper@linkstart##1##2{}\let\hyper@linkend\@empty\citealt[#1][#2]{#3}}}
\newcommandtwoopt{\citeyearads}[3][][]{\href{http://adsabs.harvard.edu/abs/#3}{\def\hyper@linkstart##1##2{}\let\hyper@linkend\@empty\citeyear[#1][#2]{#3}}}
\newcommandtwoopt{\citeadsstar}[3][][]{\href{http://adsabs.harvard.edu/abs/#3}{\def\hyper@linkstart##1##2{}\let\hyper@linkend\@empty\citealp*[#1][#2]{#3}}}
\newcommandtwoopt{\citepadsstar}[3][][]{\href{http://adsabs.harvard.edu/abs/#3}{\def\hyper@linkstart##1##2{}\let\hyper@linkend\@empty\citep*[#1][#2]{#3}}}
\newcommandtwoopt{\citetadsstar}[3][][]{\href{http://adsabs.harvard.edu/abs/#3}{\def\hyper@linkstart##1##2{}\let\hyper@linkend\@empty\citet*[#1][#2]{#3}}}
\newcommandtwoopt{\citeyearadsstar}[3][][]{\href{http://adsabs.harvard.edu/abs/#3}{\def\hyper@linkstart##1##2{}\let\hyper@linkend\@empty\citeyear*[#1][#2]{#3}}}
\newcommandtwoopt{\citeauthoradsstar}[3][][]{\href{http://adsabs.harvard.edu/abs/#3}{\def\hyper@linkstart##1##2{}\let\hyper@linkend\@empty\citeauthor*[#1][#2]{#3}}}
\newcommandtwoopt{\citepthesis}[3][][]{\href{http://tel.archives-ouvertes.fr/docs/#3}{\def\hyper@linkstart##1##2{}\let\hyper@linkend\@empty\citep[#1][#2]{#3}}}
\newcommandtwoopt{\citetthesis}[3][][]{\href{http://tel.archives-ouvertes.fr/docs/#3}{\def\hyper@linkstart##1##2{}\let\hyper@linkend\@empty\citet[#1][#2]{#3}}}
\makeatother

\newcommand{\beq}{\begin{equation}}
\newcommand{\eeq}{\end{equation}}

\newcommand{\remove}[1]{}

\journal{Computer Physics Communications}

\begin{document}

\begin{frontmatter}



\title{\usine{}: semi-analytical models for Galactic cosmic-ray propagation}

\author[label1]{David Maurin}
\ead{dmaurin@lpsc.in2p3.fr}

\address[label1]{LPSC, Universit\'e Grenoble-Alpes, CNRS/IN2P3, 53 avenue des Martyrs, 38026 Grenoble, France}

\begin{abstract}
I present the first public releases (v3.4 and v3.5) of the \usine{} code for cosmic-ray propagation in the Galaxy (\url{https://lpsc.in2p3.fr/usine}). It contains several semi-analytical propagation models previously used in the literature (leaky-box model, 2-zone 1D and 2D diffusion models) for the calculation of nuclei ($Z=1-30$), anti-protons, and anti-deuterons. For minimisations, the geometry, transport, and source parameters of all models can be enabled as free parameters, whereas nuisance parameters are enabled on solar modulation levels, cross sections (inelastic and production), and systematics of the CR data. With a single \ascii{} initialisation file to configure runs, its many displays, and the speed associated to semi-analytical approaches, \usine{} should be a useful tool for beginners, but also for experts to perform statistical analyses of high-precision cosmic-ray data.
\end{abstract}

\end{frontmatter}
%
%
{\bf PROGRAM SUMMARY}

\begin{small}
\noindent
{\em Program Title:} \usine{}.\\
{\em Journal Reference:}\\
{\em Catalogue identifier:}\\
{\em Licensing provisions:} none.\\
{\em Programming language:} C++\\
{\em Computer:} PC and Mac.\\
{\em Operating system:} UNIX(Linux), MacOS X.\\
{\em RAM:} $\sim$200-300 Mb (depends on run option and model configuration).
{\em Classification:} 1.1 Cosmic Rays.\\
{\em External routines/libraries:} \rootcern{}.\\
{\em Nature of problem:} Semi-analytical models for GCR transport.\\
{\em Solution method:} Partial derivatives on spatial coordinates solved analytically, partial derivatives on energy coordinates solved numerically (finite difference).
   \\
{\em Restrictions:} Electronic capture decay not fully implemented.
   \\
{\em Running time:} a couple of models per second (depends on model, number of energy bins, and number of cosmic rays).

\end{small}



\section{Introduction and context\label{sec:intro}}

The transport of Galactic cosmic rays (GCRs) from their sources to the Solar neighbourhood depends on many ingredients. Charged particles travel through and interact with the radiation field, the plasma, and the gas in the Galaxy. The propagation is formally described by a set of coupled MHD equations tying the CRs, gas, and magnetic field together. However, despite the impressive advances made in numerical computation in the last decade, this `first principle' approach remains challenging. For the practical purpose of interpreting CR data, a phenomenological diffusion/convection approach developed half a century ago \cite{1969ocr..book.....G} is still in used for the recent high statistic PAMELA  \cite{2014PhR...544..323A} and AMS-02 \cite[e.g.,][]{2017PhRvL.119y1101A} data.

The phenomenological approach itself is not entirely devoid of difficulties, as it amounts to solve a second order differential equation in space and momentum (see Sect.~\ref{sec:transport}) with possibly non-trivial spatial dependences of the transport coefficients and geometries. Moreover, one needs to solve a triangular matrix of coupled equations since for a given nucleus, all heavier nuclei contribute to its source term. Several approaches have been explored to solve these equations.

\paragraph{Weighted-slab and leaky-box (LB) models} In the weighted slab approach \cite{1969ocr..book.....G}, the problem is reduced to that of solving a set of coupled first-order differential equations on the path length $X$, followed by a weighted integration on the path-length distribution ${\rm PLD}(X)$. This separates fragmentation effects from other transport processes, the PLD being closely related to the associated geometry and distribution of sources \cite[e.g.,][]{1979Ap&SS..63..279L}. The even simpler LB model \cite{1960ICRC....3..220D} is a special case of the weighted slab approach with ${\rm PLD}(X)=\exp(-X/\lambda_{\rm esc}(E))$, where $\lambda_{\rm esc}(E)$ is the escape length. These models rely on the fact that no matter how complicated the geometry and spatial dependence of the ingredients are, they amount to an effective grammage crossed by the CRs, or equivalently, to an escape probability from (or confinement time in) the geometry. The leakage lifetime approximation was formally established by \cite{1970PhRvD...2.2787J}, showing at the same time that this LB approach fails and should not be used for high energy leptons, while \cite{1975Ap&SS..32..265P} showed that it also fails for low-energy radioactive species. Further complications and limitations of the weighted slab approach, even for stable species, were highlighted in \cite{1996ApJ...465..972P}. For these reasons, the weighted-slab model has been slowly abandoned, while the LB model, valid for stable species, has become less used.

\paragraph{Finite difference schemes}
With the growing wealth of CR data on nuclei, anti-nuclei, and leptons, there was a pressing need in the 90's to describe all the species in a unified framework. Solving the diffusion equation came the public code GALPROP \cite{1998ApJ...509..212S}, relying on finite difference numerical scheme \cite{1947PCPS...43...50C}; GALPROP rapidly became the new reference in the field.  This approach allows a more realistic description of the Galaxy: there is a unique system of algebraic equations per coordinate system, which applies to any spatial and energy dependence of the Galaxy parameters (transport, gas, and sources). However, the difficulty lies in choosing an optimal solver to ensure a reasonable speed and numerical stability of the solution. More recently, two other public codes based on optimised or different solvers (also based on finite differences) were released, namely DRAGON \cite{2008JCAP...10..018E,2017JCAP...02..015E} and PICARD \cite{2014APh....55...37K}. These next generation codes have been specifically designed to tackle difficult questions, such as the impact of 3D spatial structures in the sources and gas, anisotropic diffusion, etc.

\paragraph{Stochastic differential equations}
The diffusion equation can also be solved by Monte Carlo random walk \cite{1943RvMP...15....1C}. It was first used for a consistent calculation of electrons and nuclei in \cite{1997AdSpR..19..817W}. However, although simple in principle, this approach is computationally demanding. It requires a large number of particles to be tracked to limit the statistical errors on the predicted quantities, probably explaining why it remained for a long time limited to a few studies \cite{2000ApJ...528..789B,2006ApJ...639..173C,2008ApJ...681.1334F,2010JApA...31...81F,2014NewA...30...32K}. A benefit of this approach is that it can track the details related to the density probability functions of CRs reaching us or leaving the sources in time and space. Owing to the simplicity and computational efficiency on modern computer architectures, solvers for stochastic differential equations are becoming widespread \cite{2017SSRv..tmp...25S} with a public code available \cite{2012CoPhC.183..530K}. This Monte Carlo approach was used in the latest release of CRPROPA3.1 \cite{2017JCAP...06..046M}, a code for extragalactic propagation, whose low-energy extension now includes galactic propagation.

\paragraph{Semi-analytical approach and the \usine{} package}
In semi-analytical approaches, simplifying assumptions are made so that standard analysis methods (Fourier transform, Bessel expansion, Green functions) can be used to solve analytically the spatial derivatives \cite[e.g.,][]{1992ApJ...390...96W} and/or the momentum derivatives \cite[e.g.,][]{1988Ap&SS.145..319L}. The main limitation of these models is that they require a simple geometry and a simple spatial distribution of the ingredients entering the calculation (convection, diffusion, energy gains and losses). Moreover, each time a new configuration is considered, the new solution must be calculated and implemented, which differs for nuclei and leptons. The main benefit of these models is that they remain much faster than any numerical model (and by definition less sensitive to numerical instabilities). They may also provide insight into the main dependences on the ingredients through their solution. Historically, this approach was the first developed. The two-zone diffusion model (thin disc and thick diffusive halo) in 1D \cite{2001ApJ...547..264J} or 2D \cite{1992ApJ...390...96W} remain useful and provide a complementary approach to analyse recent data on nuclei and anti-nuclei \cite{2009PhRvL.102g1301D,2010A&A...516A..66P,2017PhRvL.119x1101G}\footnote{For instance, these models were used to identify the basin of sources that contribute to the measured CR fluxes \cite{2003A&A...402..971T,2003A&A...404..949M}, as was later studied by means of stochastic differential equations \cite{2010JApA...31...81F}. They are also still in use to study the stochasticity of sources \cite{1979ApJ...229..424L,2004ApJ...609..173T} and the variance on CR fluxes \cite{2012A&A...544A..92B,2017A&A...600A..68G}.}.

The \usine{} package implements several semi-analytical models. The present version has a simple text-user interface, and running or fitting CR data with \usine{} depends on a single initialisation \ascii{} file (no code recompilation needed). It also provides many pop-up and comparison plots, making it a useful tool for both beginners and experts in the field. The code and its documentation are distributed through {\sc gitlab} (see below).

The paper is organised as follows: Sect~\ref{sec:transport} recalls the master equation for Galactic CR propagation. The code is described in Sect.~\ref{sec:code} and the input ingredients in Sect.~\ref{sec:input_files}. All \usine{} runs rely on an \ascii{} initialisation file, which is described in Sect.~\ref{sec:init_file}, and run examples are provided in Sect.~\ref{sec:run_examples}. I conclude and briefly comment on the undergoing developments for the next release in Sect.~\ref{sec:conclusion}. This article describes \usine{} v3.4, used in \cite{2016PhRvL.117w1102A}, but it also highlights in \ref{app:v3.5} the improved v3.5, used for AMS-02 data analyses in \cite{2019arXiv190408210D,2019arXiv190408917G,2019arXiv190607119B}.


\section{Transport equation \label{sec:transport}}
Cosmic-ray propagation in the Galaxy can be described by a diffusion equation \cite{1990acr..book.....B,2002cra..book.....S,2007ARNPS..57..285S}:
\begin{eqnarray}
\label{eq:diffusion}
 \!\!\!\!\! \!\! \frac{\partial n(\vec r, p, t)}{\partial t} &+& \vec\nabla \cdot ( -K\vec\nabla n + \vec{V}_c n ) \\
  &+& \frac{\partial}{\partial p} \left[\dot{p} n
      + \frac{p}{3} \, (\vec\nabla \cdot \vec{V}_c )n\right]
      - \frac{\partial}{\partial p}\, p^2 K_{pp} \frac{\partial}{\partial p}\, \frac{1}{p^2}\, n  \nonumber \\
  &=& {\rm Source}(\vec r, p, t) - {\rm Sink}(\vec r, p, t)\nonumber,
\end{eqnarray}
\noindent
with $n (\vec r,p,t)$ the CR density per particle momentum $p$, $K$ the spatial diffusion coefficient, $\vec{V}_c$ the convection velocity, $\dot{p}\equiv dp/dt$ the momentum losses, and $K_{pp}$ the diffusion coefficient in momentum space for reacceleration. This equation is formally the continuity equation (first line) with the energy current (second line), and source and sink terms (last line) detailed below.

\paragraph{Source terms}
\begin{itemize}

  \item {\em Primary origin} Diffusive shock acceleration (e.g., in supernova remnants) is the favoured mechanism to accelerate all species, with an injection spectrum $\propto R^{-\alpha}$ with $\alpha\approx 2$ ($R=pc/Ze$ is the rigidity). The CRs accelerated in sources are denoted {\em primaries} for short (e.g., $^1$H, $^{16}$O, $^{30}$Si\dots).

  \item {\em Secondary origin} Nuclear interactions of primary CRs on the interstellar medium (ISM) give rise to secondary particles ($e^+$, $e^-$, $\gamma$, $\nu$), anti-nuclei ($\bar{p}$, $\bar{d}$, \dots), and secondary nuclei as fragments of heavier CR nuclei. These CRs are denoted {\em secondaries} for short (e.g., Li to B isotopes).

  \item {\em Tertiary origin} Non-annihilating inelastic nuclear interactions are reactions in which the particles survive the interactions, but loose some of their energy (e.g., in resonances). This effect was first accounted for in $\bar{p}$ by \cite{1983ApJ...269..751T}. This reaction actually both provides a net loss of particles at the energy of interaction, and a gain at lower energies coming from the redistribution of the particles in energy. These redistributed CRs are often denoted {\em tertiaries} for short, as they involve re-interactions of secondary CRs.

  \item {\em Radioactive origin} Unstable CRs are net sources for their progeny\footnote{Primary or secondary unstable/daughter species may be used as {\em cosmic-ray clocks} for the confinement time in the Galaxy \cite[e.g.,][]{2002A&A...381..539D}, {\em acceleration clocks} for the time elapsed between synthesis and acceleration \cite[e.g.,][]{2016Sci...352..677B}, or {\em reacceleration-meter} to estimate the amount of re-acceleration gained during propagation \cite[e.g.,][]{1998A&A...336L..61S}.}. The CR fraction decaying is set by the competition between decay times (e.g., 1.387~Myr for $^{10}$Be) and escape time from the Galaxy (tens of Myrs at a few GeV/n)---for instance, 15\% of $^{10}$B at 10 GeV/n comes the $\beta$-decay of secondary $^{10}$Be \cite{2018PhRvC..98c4611G}. Another decay type is electronic capture (EC) decay, in which CR ions must attach first an electron \cite[see][for a review]{1984ApJS...56..369L}, the effective half-life being a competition between electron attachment, stripping, and EC-decay \citepthesis[e.g.,][]{00/04/78/51/PDF/tel-00008773.pdf}.

\end{itemize}

\paragraph{Sink terms}
\begin{itemize}
  \item {\em Destruction} Inelastic interactions on the ISM destroy CRs. The interaction time is constant above GeV/n energies, but gradually becomes a sub-dominant process as the escape time steadily decreases with energy. Also, heavy nuclei are more prone to destruction than lighter ones, as cross sections typically scale as $A^{2/3}$.

  \item {\em Redistribution} A fraction of anti-nuclei which suffer non-annihilating interactions disappears. This is this very fraction that feeds the tertiary source term described above.

  \item {\em Decay} When unstable species decay, they change their nature. The various decay channels are the ones mirroring those described above in the radioactive source term.
\end{itemize}

All models in this release are steady-state models, including the LB model, and the 1D and 2D 2-zone (thin disc, thick halo) diffusion models. The latter two only allow for a constant convective wind perpendicular to the disc, isotropic and spatial-independent diffusion coefficients, constant gas density in the thin disc, and sources located in the thin disc only. I do not repeat the equations and their solutions as they are already detailed in \cite{2009A&A...497..991P} for the LB model, and \cite{2010A&A...516A..66P} for the 1D and 2D models. In practice, for all \usine{} models, Eq~(\ref{eq:diffusion}) is taken per unit of kinetic energy per nucleon $E_{k/n}$\footnote{This quantity is conserved for the production of nuclei in  the straight-ahead approximation.}. We have $(dn/dE_{k/n})=(A/\beta)\times(dn/dp)$ and the CR density is related to CR fluxes by $\psi = v/(4\pi) \times n$. As experiments provide their data in different energy units $E_u$, \usine{} converts them in (m$^{-2}$~s$^{-1}$~sr$^{-1}~E_u^{-1}$), with $E_u$ the rigidity $R$ (GV), the total $E$ or kinetic $E_k$ energy (GeV), or the kinetic energy per nucleon $E_{k/n}$ (GeV/n). The domain of validity of the current \usine{} models is from tens of MeV/n to tens of PeV/n (maximum energy of Galactic sources). The initialisation files (see next) in this release allow to calculate anti-deuterons, antiprotons, all stable and $\beta$-unstable CR nuclei from $Z=1-30$, but not EC-unstable secondary species; the latter require a specific solution which is not not implemented yet.

\section{Description of the code\label{sec:code}}

The code \usine{} is written in C++ (with classes). It relies on the \rootcern{}\footnote{\url{http://root.cern.ch}} library and wrappers for displays and minimisation routine. The code's structure is standard, with headers in {\tt include/*.h}, sources in {\tt src/*.cc}, compiled libraries in {\tt lib/}, objects in {\tt obj/}, and one executable in {\tt bin/}. The other directories present in the package are:

\begin{itemize}
  \item {\tt FunctionParser/} third-party library\footnote{\url{http://warp.povusers.org/FunctionParser/fparser.html}} to handle formulae for many \usine{} parameters (diffusion coefficient, source spectrum, etc.);
  \item {\tt inputs/} necessary \ascii{} files to initialise and run \usine{} (see Sect.~\ref{sec:input_files});
  \item {\tt tests/} reference test files to check that the installation is successful and \usine{} ready to run, based on both unit and integration tests;
  \item {\tt doc/} \usine{} documentation ({\sc sphinx}\footnote{\url{www.sphinx-doc.org}} with {\sc readthedocs}\footnote{\url{https://readthedocs.org}} theme).
\end{itemize}

The \usine{} package and its documentation are available at \url{https://lpsc.in2p3.fr/usine}. The compilation relies on {\sc cmake}\footnote{\url{https://cmake.org}} (files {\tt CmakeLists.txt} and {\tt FindROOT.cmake}), see the documentation for details. During the development, both static ({\sc cppcheck}\footnote{\url{http://cppcheck.sourceforge.net/}}) and dynamic ({\sc valgrind\footnote{\url{http://valgrind.org/}}}) code analyses were performed to track and fix code errors, memory leaks, etc.

\begin{figure*}[th]
\centering
\includegraphics[width=0.85\textwidth]{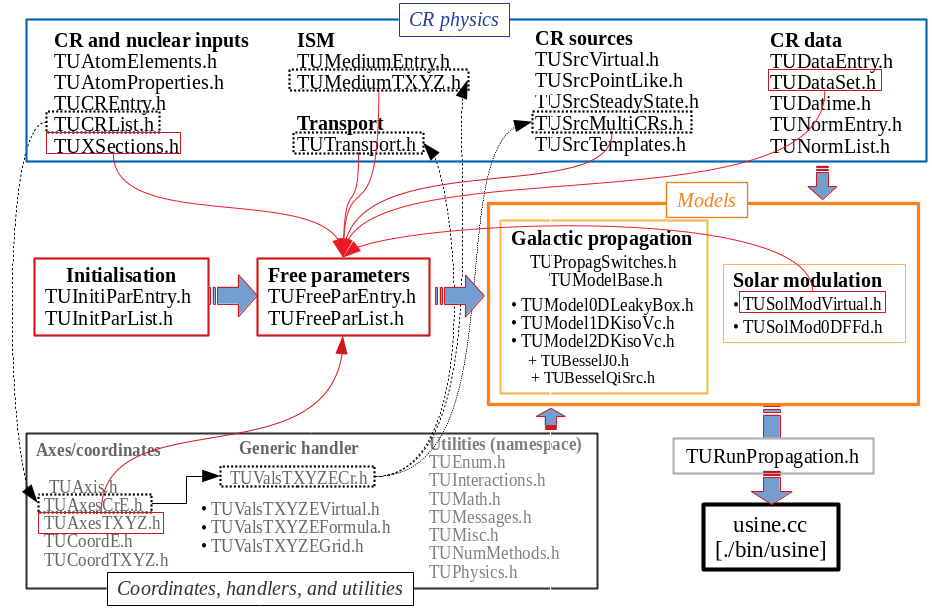}
\caption{Organisation chart of \usine{} classes.}
\label{fig:classes}
\end{figure*}
The present version of the \usine{} code was recently used to hint at the presence of a break in the diffusion coefficient from recent AMS-02 B/C data \cite{2017PhRvL.119x1101G}. It was also used to rank the most important cross sections involved in Li, Be, B, C, and N fluxes for CR studies \cite{2018PhRvC..98c4611G}. These two studies can be easily repeated and extended with the \usine{} package as delivered, as illustrated in Sect.~\ref{sec:run_examples}.

\paragraph{About previous unreleased versions}
The first \usine{} version was developed last century. With a growing interest for semi-analytical models, especially in the context of dark matter indirect detection, a public release was planned in the early 00's, but was delayed for all these years\footnote{The best advice I could provide to young researchers is not to wait for a perfect, nice, and tidy code version before making it public, because it is a never-ending process in which boredom often wins over completion. More importantly, as published science and story telling is never about delay or failure, I recommend reading \citet{kahneman2011thinking} --- Nobel Memorial Prize in Economic Sciences (2002) for his work on the psychology of judgement and decision-making, as well as behavioural economics. To quote him on one of the many biases (on decision making) human beings are prone to: {\em We focus on our goal, anchor on our plan, and neglect relevant base rates, exposing ourselves to the planning fallacy. We focus on what we want to do and can do, neglecting the plans and skills of others. Both in explaining the past and in predicting the future, we focus on the causal role of skill and neglect the role of luck. We are therefore prone to an illusion of control. We focus on what we know and neglect what we do not know, which makes us overly confident in our beliefs.}}: the first unreleased \usine{} version (in C) was used to implement for the first time automated scan of the propagation parameter space \cite{2001ApJ...555..585M,2002A&A...394.1039M} and to study the impact of the local under-density on radioactive cosmic-rays \cite{2002A&A...381..539D}; the second unreleased version (in C/C++, interfaced with \rootcern{}) included anti-protons and anti-deuterons \cite{2008PhRvD..78d3506D,2009PhRvL.102g1301D}, and was also interfaced, for the first time in the field, with a Markov Chain Monte Carlo engine \cite{2009A&A...497..991P,2010A&A...516A..66P,2011A&A...526A.101P,2012A&A...539A..88C} to extract probability density functions of the fit parameters.

\subsection{Code structure}
Figure~\ref{fig:classes} shows a sketch of USINE classes grouped by items, underlying the most relevant features of the code. A detailed description of class members and their role is provided in the header files ({\tt include/TU*.h}):
\begin{description}
   \item[CR physics] (blue box, top): classes for CR base ingredients, with CR charts, data, and cross-sections set from input files (see Sect.~\ref{sec:input_files}).
   \item[Coordinates, handlers, and utilities] (grey box, bottom): one of the most important class in \usine{} is {\tt TUValsTXYZCrE.h}, a handler for formulae (or values on a grid), dependent on generic CR, energy, and space-time coordinates. The dashed arrows indicate the classes involved and where {\tt TUValsTXYZCrE.h} is used (for transport, ISM description, and source parameters).
   \item[Models] (orange box, right-hand side): dedicated classes for Galactic propagation and Solar modulation models.
   \item[Initialisation] (red box, left-hand side): class reading the parameter file (see Sect.~\ref{sec:init_file}) to set and initialise all classes for a run.
   \item[Free parameters] (red box, centre): class handling all fit-able parameters. Red arrows connecting red boxes highlight classes with free parameters (spatial coordinates, cross sections, ISM, transport, sources, Solar modulation), see Sect.~\ref{sec:fit} for the syntax.
\end{description}

\begin{figure*}[t]
\centering
\includegraphics[width=0.8\textwidth]{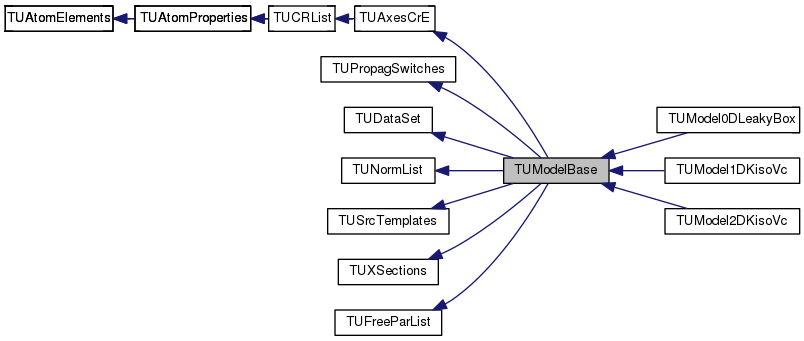}\vspace{1.5cm}\\
\includegraphics[width=0.77\textwidth]{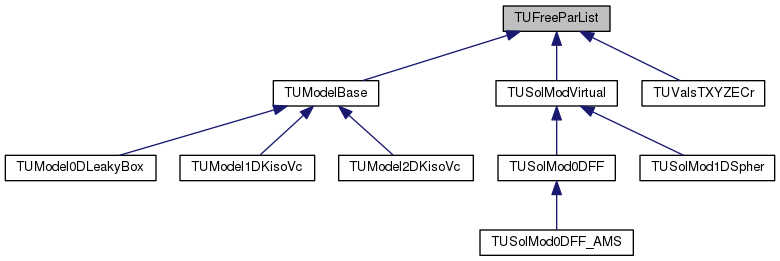}
\caption{Inheritance diagram in \usine{} for propagation model classes (top) and free parameters (bottom).}
\label{fig:inheritance}
\end{figure*}

\subsection{Inheritance diagrams}
Two important classes in \usine{} largely rely on inheritance and are shown in Fig.~\ref{fig:inheritance}, as obtained running {\sc doxygen}\footnote{\url{http://www.stack.nl/~dimitri/doxygen/}} with {\sc graphviz}\footnote{\url{http://graphviz.org}} enabled\footnote{A {\sc doxygen} documentation for developers is provided, see \usine{} webpages.}.
First, all propagation models inherit from {\tt TUModelBase.h} (top panel). The latter class gathers all ingredients shared by any model: it {\em inherits from} CR list (charts/properties) and energy ‘axis’, propagation switches (physics effects switched on or off), CR data, list of data on which to normalise primary fluxes, CR source templates (to be used as CR sources), cross-sections (inelastic, production, etc.), and a list of free parameters for the model; it also has, {\em as members}, transport parameters, ISM description, source description, and a list of all free parameters (model and ingredients).
Second, free parameters for all relevant classes are stored in a {\tt TUFreeParList.h} object. They are themselves collected in a global list of free parameters (bottom panel). This allows a very simple declaration of the parameters the user wishes to let free in a minimisation (see Sect.~\ref{sec:fit}).

\subsection{Focus on selected functions\label{sec:functions}}

I highlight in this section several important general-purpose functions, i.e. independent of the propagation model selected. All these functions belong to one of the {\tt c++} class listed in Fig.~\ref{fig:classes}, and their identifiers are thus {\tt class::function}.

\paragraph{Energy losses: {\tt TUInteractions::DEdtIonCoulomb()}}
Function to calculate Coulomb and ionisation losses in the ISM.

\paragraph{Secondaries: {\tt TUXSections::SecondaryProduction()}}
Integrates on all incoming projectile ($P$) energies, $E_{k/n}^{\rm in}$, the differential production cross sections of $F$ at $E_{k/n}^{\rm out}$:
\begin{eqnarray}
\!\!\!\!\!\frac{dQ_{\rm sec}^{P\rightarrow F}}{dE_{k/n}^{\rm out}}(E_{k/n}^{\rm out})\!\!&=&\!\!\int_{0}^{\infty} 
dE_{k/n}^{\rm in} \;\frac{dN^P}{dE_{k/n}^{\rm in}}(E_{k/n}^{\rm in}) \\
\!\!&\times&\!\! n_{\rm ISM} \times v^{\rm in} \times A_F \times \frac{d\sigma^{P+{\rm ISM} \rightarrow F}}{dE_{\rm tot}^{\rm out}}. \nonumber
\end{eqnarray}
The extra factor $A_F$ originates from the conversion from differential kinetic energy per nucleon to differential total energy, which is the format of production cross sections files (see Sect.~\ref{sec:XS}).

\paragraph{Tertiaries: {\tt TUXSections::TertiaryProduction()}}
\begin{eqnarray}
\!\!\!\!\!\frac{dQ_{\rm ter}^P}{dE_{k/n}^{\rm out}}(E_{k/n}^{\rm out}) \!\!\!&=&\!\!\! \int_{E_{k/n}^{\rm out}}^{\infty} \frac{dN^P}{dE_{k/n}^{\rm in}}(E_{k/n}^{\rm in}) n_{\rm ISM} v^{\rm out} A_P \frac{d\sigma^{P+{\rm ISM} \rightarrow P}}{dE_k^{\rm out}} dE_{k/n}^{\rm in} \nonumber\\
&-&\!\! \frac{dN^P}{dE_{k/n}^{\rm out}}(E_{k/n}^{\rm out}) \times n_{\rm ISM} \times v^{\rm out} \times \sigma^{\rm ina} (E_{k/n}^{\rm out}),
\end{eqnarray}
with $\sigma^{\rm ina}$ the inelastic non-annihilating cross section. The extra factor $A_P$ originates from the conversion from total to kinetic energy per nucleon in the differential cross section. This equation is solved iteratively \cite[e.g.,][]{2001ApJ...563..172D}, as implemented in each model class---for instance {\tt TUModel1DKisoVc::IterateTertiariesN0()}. In \usine{}, the tertiary contribution can be selected for any user-specified CR species (see Table~\ref{tab:base}), as long as the associated cross-section files exist. However, in practice, this contribution is only implemented where it matters, i.e. for anti-nuclei.

\paragraph{Solver for energy: {\tt TUModelBase::InvertForELossGain()}}
Fluxes are mostly power laws and kinetic energy per nucleon is approximately conserved in nuclear fragmentation reactions (straight-ahead approximation). This is the motivation to solve the second order differential equation in energy, Eq.~(\ref{eq:diffusion}), on a logarithmic scale in kinetic energy per nucleon. More specifically, for the model considered, one can always rewrite, after solving for the spatial coordinates (I omit the index $j$ and the $E_{k/n}$ dependence for readability):
\begin{equation}
 n + A \frac{d}{d\ln E_{k/n}} \left( B n - C \frac{dn}{d\ln E_{k/n}} \right) = S\nonumber,
 \end{equation}
with $A, B, C$ and $S$ terms depending on the model parameters, energy losses and gains, and source and sink terms. Using a finite difference scheme with boundary conditions (see {\tt gENUM\_BCTYPE} in Table~\ref{tab:enum}), this amounts to a tridiagonal matrix inversion that is inverted in {\tt TUNumMethods::SolveEq\_1D2ndOrder\_Explicit()}. A detailed description of the chosen boundary conditions, their associated coefficients in the matrix, and the impact on the solution, as well as the stability of the numerical scheme, is provided in Sect.~3.1, App.C, and App~D of \cite{2019arXiv190408210D} respectively.

\paragraph{Flux calculation: {\tt TUValsTXYZECr::OrphanVals()}} The arguments of the function are {\em quantity} (e.g., B/C, O, etc.), {\em energy type} and {\em grid} ({\tt kEKN}, {\tt kR}\dots) on which to calculate {\em quantity} (see {\tt gENUM\_ETYPE} in Table~\ref{tab:enum}), {\em position} in the model, and {\em modulation} model and level. The function proceeds as follow: (i) identify all isotopes in quantity, (ii) calculate IS flux for all isotopes at desired position, (iii) modulate all isotopes on same kinetic per nucleon grid, (iv) convert all results on desired grid and type (e.g., {\tt kR} via a log-log interpolation), and finally (v) combine all the isotopic fluxes in the desired quantity.

\paragraph{$\chi^2$ calculation: {\tt TURunPropagation::Chi2\_TOAFluxes()}} The default configuration for the $\chi^2$ calculation is to loop over all time periods $t$ (corresponding to different modulation levels), all energy types $e$ selected of all quantities $q$ selected:
\begin{equation}
  \chi^2_{\rm base} = \sum_{t, e,\, q} \left(\sum_{k=0}^{n(t,e,q)} \frac{({\rm data}_k-{\rm model}_k)^2}{\sigma_k^2}\right),
  \label{eq:chi2}
\end{equation}
with $n(t,e,q)$ the number of data and $\sigma_k$ the error on data. This calculation is modified for the following cases:
\begin{itemize}
   \item {\em Asymmetric error bars:} I rely on the standard procedure, i.e. use in Eq.~(\ref{eq:chi2}) the upper error $\sigma^{\rm up}$ (resp. $\sigma^{\rm lo}$) if the model is above the data (resp. below the data).

   \item {\em Energy bin sizes:} experiments provide the isotropic flux in the $i-$th energy bin, that is the number of events divided by the bin size. Rather than use the model value at the data mean energy, a more accurate calculation is to compare the data to
   \[
      {\rm model}_i = \frac{\int_{E_i}^{E_{i+1}} {\rm model}(E)dE}{(E_{i+1} - E_i)}.
   \]
   I assume that we have a power-law in the bin, $J=a\times E^b$, so that
   \[{\rm model}_i \approx \frac {J_{i+1}\times E_{i+1}-J_{i}\times E_{i}}{(b+1)\times E_{i+1} - E_i},
   \]
   with $b = \ln(J_{i+1}/J_i)/\ln(E_{i+1}/E_i)$.
   This assumption can be refined by adding intermediate points (on which the power-law approximation is also assumed) : this is controlled by the parameter {\tt fNExtraInBinRange} (see Table~\ref{tab:fit})\footnote{For a combo (e.g., B/C), the above calculation is performed for all isotopes separately before forming the quantity as they may have different energy dependences.}.

   \item {\em Covariance:} if a covariance matrix is provided (see Sect.~\ref{sec:crdata}) for a given $(t,e,q)$ dataset of $N_E$ energies, the corresponding term in Eq.~(\ref{eq:chi2}) is replaced with:
   \begin{equation}
      \chi^2_{\rm cov} = \sum_{\alpha} \sum_{i,j=0}^{n_E,n_E} ({\rm data}_i-{\rm model}_i) ({\cal C}^\alpha)^{-1}_{ij} ({\rm data}_j-{\rm model}_j),
      \label{eq:cov}
   \end{equation}
   with ${\cal C}^\alpha_{ij}$ the covariance matrix provided by the user for the systematics $\alpha$ (several systematics with a different covariance may exist for a given set of data).

   \item {\em Nuisance parameters:} this version enables Gaussian-distributed nuisance parameters (see Sect.~\ref{sec:fit}), each parameter ${\cal N}$ adding a contribution:
   \begin{equation}
      \chi^2_{\rm nuis} = (\frac{{\cal N}-{\cal N}_0}{\sigma_{\cal N}})^2.
      \label{eq:chi2_nuis}
   \end{equation}
   Note that if a systematic error is set as nuisance parameter, it amounts to a model bias. For $N_s$ types of systematics set as nuisance for a given quantity, the model calculation reads
   \begin{eqnarray}
       \!\!\!{\rm model}_k \!\! &=& \!\!{\rm model}^{\rm true}_k \times {\rm bias}_k\\
       &=&\!\! {\rm model}^{\rm true}_k \times \Pi_{\alpha=0\dots N_s} (1+\nu^\alpha \times \sigma^\alpha_k),\nonumber
   \end{eqnarray}
   with $\nu^\alpha$ the error nuisance parameters for systematics $\alpha$ (centred on 0, variance 1), and $\sigma^\alpha_k$ the relative error for the energy $k$ read from the diagonal of the covariance matrix (we do not need the off-diagonal terms of the covariance matrix in that case, but I still use it as it was easier to implement it this way). In this very specific case, the systematics is not added in $\chi^2_{\rm cov}$, but only as a standard $\chi^2_{\rm nuis}$ (see above). The bias is accounted for and displayed in the result plots after minimisation (see Sect.~\ref{sec:option_m}).

\end{itemize}

\subsection{Enumerators and keywords}

The code contains many enumerators ({\tt enum} key), most of them being relevant for developers only. The ones relevant for users are gathered in Table~\ref{tab:enum}. They are used for  parameter values in the initialisation file (Sect.~\ref{sec:init_file}) and/or as arguments of the executable {\tt ./bin/usine} (Sect.~\ref{sec:run_examples}). The complete list of enumerators is defined in {\tt include/TUEnum.h}.

\begin{table}[!th]
\caption{Important \usine{} enumerators and associated keywords.}
\label{tab:enum}
\begin{center}
{
\footnotesize
\begin{tabular}{lll}  \hline
\!\!\!Enumerators                 &  Keywords                 & Description                          \\\hline
\!\!\!{\tt gENUM\_ETYPE}$^\star$  & {\tt kEKN}                & Kinetic E per nuc [GeV/n]\!\!\!      \\[-0.05cm]
                                  & {\tt kR}                  & Rigidity [GV]                        \\[-0.05cm]
                                  & {\tt kETOT}               & Total energy [GeV]                   \\[-0.05cm]
                                  & {\tt kEK}                 & Kinetic energy [GeV]                 \\[+0.1cm]
\!\!\!{\tt gENUM\_ERRTYPE}$^\S$   & {\tt kERRSTAT}            & Statistical only                     \\[-0.05cm]
                                  & {\tt kERRSYST}            & Systematic only                      \\[-0.05cm]
                                  & {\tt kERRTOT}             & Stat.+ syst. in quadrature\!\!\!     \\[-0.05cm]
                                  & {\tt kERRCOV}             & Covariance matrix of errors\!\!\!\!  \\[+0.1cm]
\!\!\!{\tt gENUM\_FCNTYPE}$^\circ$& {\tt kFORMULA}            & Formula (equation)                   \\[-0.05cm]
                                  & {\tt kGRID}               & Values on a grid                     \\[+0.1cm]
\!\!\!{\tt gENUM\_BCTYPE}$^\dagger$& {\tt kNOCHANGE}          &  Solution w/o = w/ E-losses\!\!\!    \\[-0.05cm]
                                  & {\tt kD2NDLNEKN2\_ZERO}\!\!\!& $\partial^2n/\partial(\ln E_{k/n})^2=0$                  \\[-0.05cm]
                                  & {\tt kNOCURRENT}          & $j_E=0$                              \\[-0.05cm]
                                  & {\tt kDFDP\_ZERO}         & Flux derivative null                 \\[+0.1cm]
\!\!\!{\tt gENUM\_PROPAG\_MODEL}\!\!\! & {\tt kMODEL0DLEAKYBOX}\!\!\!\!\!& Leaky-box model           \\[-0.05cm]
                                  & {\tt kMODEL1DKISOVC}\!\!\!& 1D diffusion model                   \\[-0.05cm]
                                  & {\tt kMODEL2DKISOVC}\!\!\!& 2D diffusion model                   \\[+0.1cm]
\!\!\!{\tt gENUM\_SOLMOD\_MODEL}\!\!& {\tt kSOLMOD0DFF}       &  Force-Field approximation\!\!\!     \\[-0.00cm]
\hline\hline
\end{tabular}
\\$^\star$ Energy type/unit for fits or displays.
\\$^\S$ Data errors for fits or displays.
\\$^\circ$ Parametrisation types for diffusion coefficient, source spectrum, etc.
\\$^\dagger$ Boundary condition at low or high energy end.
\\
}
\end{center}
\vspace{-.5cm}
\end{table}


\section{Input \ascii{} files (ingredients)\label{sec:input_files}}

Galactic cosmic-ray propagation requires many inputs, which are handled by as many classes in \usine{}, as shown in Fig.~\ref{fig:classes}. Input files are \ascii{} files and default choices are given in {\tt inputs/}. Users can feed to \usine{} their own files and can locate them anywhere, as long as the path to the file is provided in the initialisation file. I give a short description below, and refer the reader to the online documentation for more details on the file format.

\subsection{Atomic properties and CR charts}

The file {\tt atomic\_properties.dat} gathers atomic properties for all possible CR elements. Some of these properties may be used for propagation, e.g. to account for an acceleration bias in source terms or for the calculation of electron attachment and stripping for electronic-capture (EC) decay.

The files {\tt crcharts\_*.dat} provide charts (mass, Z, A\dots) for CRs, and we only need to consider stable nuclei and unstable whose half-life is not too small compared to the propagation time\footnote{There is a subtlety for EC-unstable nuclei as their decay time may be tiny while their effective half-life, driven by electron attachment time (CRs are fully ionised above GeV/n energies), may be of the order of the propagation time. This is an issue mostly for heavy nuclei, as many of them are EC unstable.}. Ghost nuclei, unused for most \usine{} runs, are also listed: they are short-lived isotopes whose decay chain leads to a stable (or long-lived) CR \cite{1984ApJS...56..369L}, and they enter the calculation of the cumulative cross-section into a given CR \cite[e.g.,][]{2018PhRvC..98c4611G}.

\subsection{Solar system (SS) abundances}

Ideally, SS abundances should be used for comparison purpose after retro-propagating CR measurements back to the sources. However, propagation runs focusing on transport parameters use fixed values for the isotopic composition of elements, because CR isotopic data are scarce and only exist at low energy. The isotopic abundances of GCR sources can be initialised from {\tt solarsystem\_abundances2003.dat}, which gathers SS abundances for long-lived and stable isotopes and elements up to Uranium.

\subsection{Cosmic-ray data and covariance matrix\label{sec:crdata}}

The files {\tt crdata\_*.dat} contains list of data points (energy, value, date, name of experiment, etc.) directly retrieved from the cosmic-ray data base, CRDB\footnote{\url{http://lpsc.in2p3.fr/crdb}} \cite{2014A&A...569A..32M}. One can also create her/his own data file from recent measurements or simulated data, and add it to the list of files to load in the initialisation file (Sect.~\ref{sec:init_file}). All energy types ({\tt gENUM\_ETYPE}) listed in Table~\ref{tab:enum} are enabled, so that CR data can be as a function of kinetic energy per nucleon, rigidity, etc.

The use of a covariance matrix of errors for minimisation is enabled in \usine{}, and it requires one covariance matrix file per measurement (e.g., in directory {\tt CRDATA\_COVARIANCE/}). Actually, for past measurements, at best, only the overall uncertainties (statistical and systematics combined) were provided. In more recent data, at least statistical and systematic uncertainties are provided separately (format of data gathered in CRDB). Recent experiments (e.g., AMS02) go further and provide also systematics broken down in various contributions. However, this may not be sufficient for the model analysis of high-precision data, which may require the error correlations between different energies for all types of errors in the instrument, as encoded in the covariance matrix. The other option to deal with systematics uncertainties is via nuisance parameters (see Sect.~\ref{sec:fit}), and it is not clear yet which is the best way to account for them.

\subsection{Cross sections\label{sec:XS}}
The directories {\tt XS\_NUCLEI/} and {\tt XS\_ANTINUC/} provide energy-dependent cross sections for nuclei and anti-nuclei respectively, for both inelastic and production cross sections (straight-ahead or differential). For nuclei, production cross section files come in two flavours: those in which reactions for `ghosts' (i.e. short-lived nuclei \cite{1984ApJS...56..369L,00/04/78/51/PDF/tel-00008773.pdf}) are explicitly provided ({\tt XS\_NUCLEI/GHOST\_SEPARATELY/}), and those with cumulative cross sections ({\tt XS\_NUCLEI/}). For instance, $^{10}{\rm B}$ has a single ghost nucleus, and its cumulative cross section from a CR species $X$ reads
  \[ 
  \sigma^{\rm cumul}_{{\rm X}\rightarrow ^{10}{\rm B}} = \sigma_{{\rm X}\rightarrow {\rm ^{10}B}} + \sigma_{{\rm X}\rightarrow {\rm ^{10}C}} \times {\cal B}r({\rm ^{10}B}\rightarrow {\rm ^{10}C})\,,
  \] 
with ${\cal B}r({\rm ^{10}B}\!\!\rightarrow\!\! {\rm ^{10}C})=100\%$. Note that the heavier the species, the more ghost there are in general, though with smaller branching ratios. Whether files from {\tt XS\_NUCLEI/} or {\tt XS\_NUCLEI/GHOST\_SEPARATELY/} are used, only stable and `long-lived' ($\gtrsim$~kyr) species are propagated in \usine{}; in the latter case, cumulative cross sections are re-calculated and weighted from the branching ratios of ghosts listed in\footnote{Loading or not ghost lists is decided at initialisation stage, via {\tt IsLoadGhosts} parameter (see Table~\ref{tab:base}).} {\tt crcharts\_*.dat} files (see online documentation). I recommend the user to rely on files with cumulative cross sections, unless separate cross sections are required for a specific use (e.g., \cite{2018PhRvC..98c4611G}). Indeed, using cumulated cross sections ensures much faster propagation calculations for the same result. Speed and flexibility are also the reasons why I preferred pre-calculated cross sections over re-calculation for each run.

The format, content, and references for \usine{} cross-section files (inelastic and production) are given in the online documentation\footnote{\url{https://dmaurin.gitlab.io/USINE/input_xs_data.html}}. Some of the production cross sections were for instance obtained running WNEW \cite[e.g.,][]{1990PhRvC..41..566W} and YIELDX \cite[e.g.,][]{1998ApJ...501..920T} codes with the \usine{}\footnote{The pieces of code to do so were present in previous unreleased \usine{} version, but are not part of this release.} or GALPROP \cite{2018PhRvC..98c4611G} list of ghosts. For more details and references on the various models behind these cross sections, we refer the reader to \cite{2018PhRvC..98c4611G} and \cite{2001ApJ...563..172D,2005PhRvD..71h3013D,2007PhRvD..75h3006B,2008PhRvD..78d3506D,2017JCAP...02..048W,2018PhRvD..97j3019K,2019arXiv190607119B} for nuclei and anti-nuclei respectively.

\section{Initialisation file, models, and parameters\label{sec:init_file}}

Any calculation with \usine{} starts by loading an \ascii{} \usine{}-formatted initialisation file. The file usage is flexible enough so that there is no need to recompile the code or go into it. The price to pay is a complicated syntax for the parameter values, which are described in a specific section of the code documentation. I only focus here on the structure and goal of the initialisation file.

The quantities initialised are the list of CRs to propagate and their parents (and associated charts), the energy ranges for the various species (nuclei, anti-nuclei, leptons), the cross sections and CR data to use, and then the propagation (ISM, source, transport) and solar modulation models. The parameters to fit, their range, and whether to use them as free or nuisance parameters is also completely handled by the same \ascii{} file.

\subsection{Initialisation file syntax}

\begin{table*}[!th]
\caption{Example of base parameters from the initialisation file (Sect.~\ref{sec:init_file}). Only the last column (Value) should be edited.}
\label{tab:base}
\begin{center}
{
\small
\begin{tabular}{lll}  \hline
Group~~~~Subgroup              & ~~~~Parameter  & \!\!\!\!\!\!\!\!\!\!Multi-entry?~~~Value\\\hline
{\tt Base} @ {\tt CRData     } & @ {\tt fCRData          } &       @ {\tt M=1} @ {\tt \$USINE/inputs/crdata\dots.dat}$^\dagger$\\
{\tt Base} @ {\tt CRData     } & @ {\tt NormList         } &       @ {\tt M=0} @ {\tt H,He:PAMELA$|$20$|$kEkn;C,N,O,F,Ne,Na,Mg,Al,Si:HEAO$|$10.6$|$kEkn}\\
{\tt Base} @ {\tt EnergyGrid } & @ {\tt NBins            } &       @ {\tt M=0} @ {\tt 129}\\
{\tt Base} @ {\tt EnergyGrid } & @ {\tt NUC\_EknRange    } &       @ {\tt M=0} @ {\tt [5.e-2,5.e3]}\\
{\tt Base} @ {\tt EnergyGrid } & @ {\tt ANTINUC\_EknRange}\!\!\! & @ {\tt M=0} @ {\tt [5e-2,1.e2]}\\
{\tt Base} @ {\tt ListOfCRs  } & @ {\tt fAtomicProperties}\!\! &   @ {\tt M=0} @ {\tt \$USINE/inputs/atomic\_properties.dat}$^\dagger$\\
{\tt Base} @ {\tt ListOfCRs  } & @ {\tt fChartsForCRs    } &       @ {\tt M=0} @ {\tt \$USINE/inputs/crcharts\dots.dat}$^\dagger$\\
{\tt Base} @ {\tt ListOfCRs  } & @ {\tt IsLoadGhosts     } &       @ {\tt M=0} @ {\tt false}\\
{\tt Base} @ {\tt ListOfCRs  } & @ {\tt ListOfCRs        } &       @ {\tt M=0} @ {\tt [2H-BAR,30Si]}\\
{\tt Base} @ {\tt ListOfCRs  } & @ {\tt ListOfParents    } &       @ {\tt M=0} @ {\tt 2H-bar:1H-bar,1H,4He;1H-bar:1H,4He}\\
{\tt Base} @ {\tt ListOfCRs  } & @ {\tt PureSecondaries  } &       @ {\tt M=0} @ {\tt Li,Be,B}\\
{\tt Base} @ {\tt ListOfCRs  } & @ {\tt SSRelativeAbund  } &       @ {\tt M=0} @ {\tt \$USINE/inputs/solarsystem\_abundances2003.dat}$^\dagger$\\
{\tt Base} @ {\tt MediumCompo} & @ {\tt Targets          } &       @ {\tt M=0} @ {\tt H,He}\\
{\tt Base} @ {\tt XSections  } & @ {\tt Tertiaries       } &       @ {\tt M=0} @ {\tt 1H-bar,2H-bar}\\
{\tt Base} @ {\tt XSections  } & @ {\tt fTotInelAnn      } &       @ {\tt M=1} @ {\tt \$USINE/inputs/XS\_NUCLEI/sigInel\dots.dat}$^\dagger$\\
{\tt Base} @ {\tt XSections  } & @ {\tt fProd            } &       @ {\tt M=1} @ {\tt \$USINE/inputs/XS\_NUCLEI/sigProd\dots.dat}$^\dagger$\\
{\tt Base} @ {\tt XSections  } & @ {\tt fProd            } &       @ {\tt M=1} @ {\tt \$USINE/inputs/XS\_ANTINUC/dSdEProd\dots.dat}$^\dagger$\\
{\tt Base} @ {\tt XSections  } & @ {\tt fTotInelNonAnn   } &       @ {\tt M=1} @ {\tt \$USINE/inputs/XS\_ANTINUC/sigInelNONANN\dots.dat}$^\dagger$\\
{\tt Base} @ {\tt XSections  } & @ {\tt fdSigdENAR       } &       @ {\tt M=1} @ {\tt \$USINE/inputs/XS\_ANTINUC/dSdENAR\dots.dat}$^\dagger$\\
\hline\hline
\end{tabular}
\\ \footnotesize
$^\dagger$ Any environment variable in file names (e.g., \$USINE) is correctly interpreted.
}
\end{center}
\vspace{-.4cm}
\end{table*}

An initialisation file consists of lines whose syntax is
\begin{tabular}{p{-0.25cm}p{8.cm}}
    & {\tt group @subgroup @parameter @M=\dots @value}\\
\end{tabular}
with many predefined keywords (see Tables~\ref{tab:base} to \ref{tab:fit}). The rationale behind this choice is the following:

\begin{description}
   \item[Group/subgroup/parameter] A parameter belongs to a subgroup, which itself belongs to a group. This allows to easily associate a parameter to a given physics ingredient and organise it in the initialisation file---the list of keywords is of course a matter of the developer’s taste, and for the user’s point of view, it is just what it is (not editable).
   \item[Multiple-entry parameter] Most of the time, parameters are single-valued ones ({\tt M=false} or 0). However, CR data and cross-section files are enable for multiple-entries ({\tt M=true} or 1), i.e. several files: in such case, the files are read sequentially, and the last read values always override previously read ones (if applies).
   \item [Value] This is editable by the user, to select files, parameter values, source spectrum energy dependences, fit parameters\dots each parameter value has a specific syntax.
\end{description}

 We stress that the information (usage, specific syntax, example values) on any \usine{} keyword can be accessed from the online documentation using the search box\footnote{\url{https://lpsc.in2p3.fr/usine}.}.

\subsection{Base parameters\label{sec:base}}

Base parameters consist in parameters that are required regardless of the propagation model used. Examples of base parameters are given in Table~\ref{tab:base}, and they correspond to:
\begin{itemize}

   \item {\tt Base@CRData}: selection of generic data sets and specific list of experiment and quantities (in the data sets) to renormalise the model to. For instance, the value {\tt H,He:PAMELA$|$20$|$kEkn;C,N,O,\dots} in Table~\ref{tab:base} means that \usine{} will search for available H and He PAMELA data closest to 20~GeV/n (and a different set for C, N, O\dots), and use these points to renormalise the associated source terms keeping relative isotopic abundances fixed (unless these source terms are explicitly set as fit parameters).

   \item {\tt Base@EnergyGrid}: different families of species (nuclei and anti-nuclei) are calculated on different kinetic energy per nucleon ranges, $[E_{\rm k/n}^{\rm min}-E_{\rm k/n}^{\rm max}]$, all on $N_{\rm bins}$ logarithmically-spaced bins:
   \begin{equation}
     E_{\rm k/n}^i = E_{\rm k/n}^{\rm min} \times \Delta^i, \quad{\rm with}\;\; \Delta = \left(\frac{E_{\rm k/n}^{\rm max}}{E_{\rm k/n}^{\rm min}}\right)^{1/(N_{\rm bins} - 1)}.
     \label{eq:log-axis}
   \end{equation}
   For anti-nuclei, integrations are involved (primary and tertiary source terms) and a doubling step integration is implemented to check the convergence on the available grid: the optimal number of bins to take is
   \begin{equation}
      N_{\rm bins} = 2^j + 1.
   \end{equation}
   Note that E-range for nuclei should be larger than that for anti-nuclei, otherwise the high-energy flux of the latter is underestimated (protons at $E_p$ create $\bar{p}$ at $E_{\bar{p}}<E_p$).

   \item {\tt Base@ListOfCRs}: (i) selection of atomic and nuclear charts files to select propagated CRs---value {\tt [2H-BAR,30Si]} in Table~\ref{tab:base} means all (anti-)nuclei from $\bar{d}$ to $^{30}$Si; (ii) selection of the list of parents producing them---value{\tt 1H-bar:1H,4He} in Table~\ref{tab:base} means that the parents for $\bar{p}$ are set to $^1$H and $^4$He only; (iii) set the list of possible pure secondary species---{\tt Li,Be,B} in Table~\ref{tab:base} means that there is no source term for all isotopes of these elements.

   \item {\tt Base@MediumCompo}: Choice of the ISM target elements for the propagation (e.g., {\tt H,He} in Table~\ref{tab:base}). Any list is possible, without any change in the code (everything is allocated dynamically), but keep in mind that the production cross sections on these elements must then be provided.

   \item {\tt Base@XSections}: Choice of the cross section files used for propagation. Internally, the code loops on all the files (all parameters are multi-entry ones). It fills and checks that all the cross sections associated to the list of CRs, parents ({\tt Base@ListOfCRs}), and ISM targets ({\tt Base@MediumCompo}) are found (inelastic and production cross-sections).
\end{itemize}

\subsection{Model parameters\label{sec:model}}

\begin{table*}[!th]
\caption{Example of model parameters from the initialisation file (Sect.~\ref{sec:init_file}). Only the last column (Value) should be edited.}
\label{tab:model}
\begin{center}
{
\small
\begin{tabular}{lll}  \hline
Group~~~~~~~~~~~~~~~~~~~~~Subgroup   & ~~~~Parameter  & \!\!\!\!\!\!\!\!\!\!Multi-entry?~~~Value\\\hline
{\tt Model1DKisoVc}\! @\! {\tt Geometry}       &@ {\tt ParNames}         &@ \!{\tt M=0}\! @ {\tt L,h,rhole}$^\dagger$\\
{\tt Model1DKisoVc}\! @\! {\tt Geometry}       &@ {\tt ParUnits}         &@ \!{\tt M=0}\! @ {\tt kpc,kpc,kpc}$^\dagger$\\
{\tt Model1DKisoVc}\! @\! {\tt Geometry}       &@ {\tt ParVals }         &@ \!{\tt M=0}\! @ {\tt 8.,0.1,0.}\\
{\tt Model1DKisoVc}\! @\! {\tt Geometry}       &@ {\tt XAxis}            &@ \!{\tt M=0}\! @ {\tt z:[0,L],10,LIN}\\
{\tt Model1DKisoVc}\! @\! {\tt Geometry}       &@ {\tt XSun}             &@ \!{\tt M=0}\! @ {\tt 0.}\\
{\tt Model1DKisoVc}\! @\! {\tt ISM}            &@ {\tt Density}          &@ \!{\tt M=1}\! @ {\tt HI:FORMULA$|$0.867}\\
{\tt Model1DKisoVc}\! @\! {\tt ISM}            &@ {\tt Density}          &@ \!{\tt M=1}\! @ {\tt \dots}\\
{\tt Model1DKisoVc}\! @\! {\tt ISM}            &@ {\tt Te}               &@ \!{\tt M=0}\! @ {\tt FORMULA$|$1.e4}\\
{\tt Model1DKisoVc}\! @\! {\tt SrcSteadyState}\!\!\!\!\!\!\!&@ {\tt Species}\!\!\!\!\!\!\!&@ \!{\tt M=1}\! @ {\tt STD$|$ALL}\\
{\tt Model1DKisoVc}\! @\! {\tt SrcSteadyState}\!\!\!\!\!\!\!&@ {\tt SpectAbundInit}\!\!\!\!\!\!\!&@ \!{\tt M=1}\! @ {\tt STD$|${\tt kSSISOTFRAC,kSSISOTABUND,kFIPBIAS}}\\
{\tt Model1DKisoVc}\! @\! {\tt SrcSteadyState}\!\!\!\!\!\!\!&@ {\tt SpectTempl}\!\!\!\!\!\!\!&@ \!{\tt M=1}\! @ {\tt STD$|$POWERLAW$|$q}\\
{\tt Model1DKisoVc}\! @\! {\tt SrcSteadyState}\!\!\!\!\!\!\!&@ {\tt SpectValsPerCR}\!\!\!\!\!\!\!&@ \!{\tt M=1}\! @ {\tt STD$|$q[{\tt PERCR:DEFAULT=}1e-5]};{\tt alpha[{\tt SHARED:}2.]};{\tt eta\_s[{\tt SHARED:}-1]}\!\!\!\!\!\!\!\!\!\!\!\!\!\!\\
{\tt Model1DKisoVc}\! @\! {\tt SrcSteadyState}\!\!\!\!\!\!\!&@ {\tt SpatialTempl}\!\!\!\!\!\!\!&@ \!{\tt M=1}\! @ {\tt STD$|$-}\\
{\tt Model1DKisoVc}\! @\! {\tt SrcSteadyState}\!\!\!\!\!\!\!&@ {\tt SpatialValsPerCR}\!\!\!\!\!\!\!&@ \!{\tt M=1}\! @ {\tt STD$|$-}\\
{\tt Model1DKisoVc}\! @\! {\tt Transport}      &@ {\tt ParNames}         &@ \!{\tt M=0}\! @ {\tt Va,Vc,K0,delta,eta\_t,Rb,Db,sb}\\
{\tt Model1DKisoVc}\! @\! {\tt Transport}      &@ {\tt ParUnits}         &@ \!{\tt M=0}\! @ {\tt km/s,km/s,kpc\^{}2/Myr,-,-,GV,-,-}\\
{\tt Model1DKisoVc}\! @\! {\tt Transport}      &@ {\tt ParVals}          &@ \!{\tt M=0}\! @ {\tt 85,19,0.035,0.6,-0.1,125.,0.2,0.01}\\
{\tt Model1DKisoVc}\! @\! {\tt Transport}      &@ {\tt Wind}             &@ \!{\tt M=1}\! @ {\tt W0:FORMULA$|$Vc}$^\dagger$\\
{\tt Model1DKisoVc}\! @\! {\tt Transport}      &@ {\tt VA}               &@ \!{\tt M=0}\! @ {\tt FORMULA$|$Va}$^\dagger$\\
{\tt Model1DKisoVc}\! @\! {\tt Transport}      &@ {\tt K}                &@ \!{\tt M=1}\! @ {\tt K00}:{\tt FORMULA}$|$beta\^{}{\tt eta\_t*K0*}Rig\^{}{\tt delta*(1+(}Rig{\tt /Rb)}\^{}{\tt (Db/sb))}\^{}{\tt (-sb)}\!\!\!\!\!\!\!\!\!\!\!\!\!\!\\
{\tt Model1DKisoVc} @ {\tt Transport}      &@ {\tt Kpp}              &@ \!{\tt M=0}\! @ {\tt FORMULA$|$(4./3.)*(Va*1.022712e-3*}beta*Etot)\^{}{\tt 2/(K00*delta*\dots)}\!\!\!\!\!\!\!\!\!\!\!\!\!\!\\
{\tt TemplSpectrum}\! @\! {\tt POWERLAW}       &@ {\tt ParNames}         &@ \!{\tt M=0}\! @ {\tt q,alpha,eta\_s}\\
{\tt TemplSpectrum}\! @\! {\tt POWERLAW}       &@ {\tt ParUnits}         &@ \!{\tt M=0}\! @ {\tt /(GeV/n/m3/Myr)$^\dagger$,-,-}\\
{\tt TemplSpectrum}\! @\! {\tt POWERLAW}       &@ {\tt Definition}       &@ \!{\tt M=0}\! @ {\tt FORMULA$|$q*beta\^{}(eta\_s)*Rig\^{}(-alpha)}\\
\hline\hline
\end{tabular}
\\ \footnotesize
$\dagger$ These values cannot be edited by the user, they are intrinsic properties of the model.
}
\end{center}
\vspace{-.4cm}
\end{table*}

Each propagation model is associated with a keyword in the initialisation file, with 3 enabled models: {\tt Model0DLeakyBox} \cite{2009A&A...497..991P}, {\tt Model1DKisoVc} \cite{2010A&A...516A..66P}, and {\tt Model2DKisoVc} \cite{2001ApJ...555..585M,2001ApJ...563..172D,2010A&A...516A..66P}. There are several mandatory subgroups for any galactic propagation models, and the syntax of their parameter values is illustrated in Table~\ref{tab:model} for the 1D model, {\tt Model1DKisoVc}. Note that geometry and ISM parameter names and units are uniquely associated to a given semi-analytical models and should never be modified.
\begin{itemize}

   \item {\tt Model1DKisoVc@Geometry}: Geometry, coordinate system, and Sun’s position for the model. Here, we have as parameters the half-halo size $L$ [kpc], the thin-disc half-size $h$ [kpc] (whose value should better not be changed), and a possible gas underdensity $rhole$ [kpc] (as described in \cite{2002A&A...381..539D,2010A&A...516A..66P}). The 1D geometry is a linear grid in $z$ (first axis {\tt Xaxis}) from 0 to $L$, and all geometry parameter names are enabled as free parameters.

   \item {\tt Model1DKisoVc@ISM}: ISM parameters for the model. In all implemented models, the gas density is constant in the thin disc, so we do not need to declare specific free parameters for the ISM, and simply specify the ISM densities (HI, HII, H2, and heavier elements)\footnote{Add as many lines as the number of elements in {\tt Base@MediumCompo}. Note that H is a special case requiring HI (atomic), HII ionised), and H2 (molecular), with $n_H\equiv n_{HI}+n_{HII}+2n_{H2}$ used to calculate cross sections on H.} and properties (e.g., $T^\circ$).

\begin{table*}[!th]
\caption{Example of model parameters from the initialisation file (Sect.~\ref{sec:init_file}). Only the last column (Value) should be edited.}
\label{tab:run}
\begin{center}
{
\small
\begin{tabular}{lll}  \hline
Group~~~~~~~~~~Subgroup            & ~~~~Parameter  & \!\!\!\!\!\!\!\!\!\!Multi-entry?~~~Value\\\hline
{\tt UsineRun}@{\tt Calculation}   & @ {\tt BC\_NUC\_LE}     & @ {\tt M=0} @ {\tt kD2NDLNEKN2\_ZERO}\\
{\tt UsineRun}@{\tt Calculation}   & @ {\tt BC\_NUC\_HE}     & @ {\tt M=0} @ {\tt kNOCHANGE}\\
{\tt UsineRun}@{\tt Calculation}   & @ {\tt EPS\_ITERCONV}   & @ {\tt M=0} @ {\tt 1.e-6}\\
{\tt UsineRun}@{\tt Calculation}   & @ {\tt EPS\_INTEGR}     & @ {\tt M=0} @ {\tt 1.e-4}\\
{\tt UsineRun}@{\tt Calculation}   & @ {\tt EPS\_NORMDATA}   & @ {\tt M=0} @ {\tt 1.e-10}\\
{\tt UsineRun}@{\tt Calculation}   & @ {\tt IsUseNormList}   & @ {\tt M=0} @ {\tt true}\\
{\tt UsineRun}@{\tt Display}       & @ {\tt QtiesExpsEType}  & @ {\tt M=0} @ {\tt ALL}\\
{\tt UsineRun}@{\tt Display}       & @ {\tt ErrType}         & @ {\tt M=0} @ {\tt kERRSTAT}\\
{\tt UsineRun}@{\tt Display}       & @ {\tt FluxPowIndex}    & @ {\tt M=0} @ {\tt 2.8}\\
{\tt UsineRun}@{\tt Models}        & @ {\tt Propagation}     & @ {\tt M=1} @ {\tt Model1DKisoVc}\\
{\tt UsineRun}@{\tt Models}        & @ {\tt SolarModulation} & @ {\tt M=1} @ {\tt SolMod0DFF}\\
{\tt UsineRun}@{\tt OnOff}         & @ {\tt IsDecayBETA}     & @ {\tt M=0} @ {\tt true}\\
{\tt UsineRun}@{\tt OnOff}         & @ {\tt IsDestruction}   & @ {\tt M=0} @ {\tt true}\\
{\tt UsineRun}@{\tt OnOff}         & @ {\tt \dots}           & @ {\tt M=0} @ {\tt true}\\
\hline\hline
\end{tabular}
}
\end{center}
\vspace{-.4cm}
\end{table*}

   \item {\tt Model1DKisoVc@SrcSteadyState}: steady-state source description (CR content, spectra, and spatial distribution if needed) based on templates for the spectra and spatial distributions. For the model shown in Table~\ref{tab:model}, I defined a spectral template, {\tt TemplSpectrum@POWERLAW}\footnote{Templates are the only cases for which the {\tt subgroup} keywords (here {\tt POWERLAW}) can be edited; the user is free to use any name.}. The syntax for formulae is that of {\sc fparser}\footnote{\url{http://warp.povusers.org/FunctionParser/fparser.html}}, and in this example, the generic spectrum is:
   \begin{equation}
      {\rm Template} =  q \times \beta^{eta\_s}\times R^{-alpha},
      \label{eq:powerlaw}
   \end{equation}
   where Rig, beta, gamma, p, Ekn, Ek, Etot are interpreted by \usine{} as the CR rigidity, $\beta$, $\gamma$, etc. The template spectrum must always have a normalisation parameter (here $q$). The syntax to define the source spectrum for a specific list of species, is (i) to pick a name for this source (here {\tt STD}); (ii) to specify which template it is based on and what is the normalisation parameter for this template (here {\tt POWERLAW} with {\tt q}); (iii) to specify how many free parameters will be created and their default value (all parameters of templates are enabled as free parameters). In this example, there is one free parameter per CR ({\tt q\_1H, q\_2H\dots q\_30Si} (internally created in \usine{}), and universal free parameters {\tt eta\_s} and {\tt alpha}. Note that for the 1D model, there is no source spatial dependence, so the associated parameter values are empty ({\tt -}).

   \item {\tt Model1DKisoVc@Transport}: \usine{} enables the most generic description for the transport parameters, i.e. diffusion tensor $K_{ij}$, wind vector $W_i$, and reacceleration ($V_a$ and $K_{pp}$). For the isotropic and homogeneous diffusion, with constant wind perpendicular to the disc, the 1D and 2D models only allow for the components {\tt K00} and {\tt W0}. As for the source spectrum, any function of (Rig, beta, gamma, p, Ekn, Ek, Etot) and user-defined names is enabled. In the example of Table~\ref{tab:model},
   \begin{equation}
      K_{00} =  \beta^{\eta_t} K_0 R^\delta \times \left[1+\left(\frac{R}{R_b}\right)^{\delta_b/s_b}\right]^{-s_b},
      \label{eq:k00}
   \end{equation}
   corresponding to a broken power-law as used in \cite{2017PhRvL.119x1101G}. Any of the transport parameters declared in {\tt Model1DKisoVc@Transport@ParNames} can be used as a fit parameter.

\end{itemize}

\subsection{Run parameters\label{sec:run}}

These parameters are related to the selection of models (propagation and solar modulation), numerical precision for propagation calculation, and quantities to show in displays, as illustrated in Table~\ref{tab:run}.

\begin{itemize}

   \item {\tt UsineRun@Calculation}: this includes boundary conditions (BC) to use on both low and high-energy ends (see keywords in Table~\ref{tab:enum}), the relative precision for integrations (e.g., secondary contribution from differential cross sections), iterative procedure (e.g, tertiaries for anti-nuclei), and normalisation to data.

   \item {\tt UsineRun@Display}: this is purely for display purpose, to select which data will be shown with which errors. Use {\tt ALL} to display all available data (see Sect.~\ref{sec:input_files}), or the syntax {\tt Qty1,Qty2:Exp1,Exp2:EType1;Qty3:Exp3:EType2}\footnote{For instance, {\tt He:AMS,BESS:kEKN;B/C:AMS:kR} means only He($E_{k/n}$) data from AMS and BESS and B/C(R) data from AMS.}. The error shown are selected from {\tt gENUM\_ERRTYPE} (see Table~\ref{tab:enum}).

   \item {\tt UsineRun@Models}: Propagation and Solar modulation models to use in run are selected from {\tt gENUM\_PROPAG\_MODEL} and {\tt gENUM\_SOLMOD\_MODEL} (see Table~\ref{tab:enum}).

   \item {\tt UsineRun@OnOff}: All physics effects---decay, inelastic interaction, energy losses, etc.---can be enabled or disabled in \usine{}. This is especially useful for comparison purpose (see Sect.~\ref{sec:run_examples}).

\end{itemize}

\subsection{Fit parameters\label{sec:fit}}

\begin{table*}[!th]
\caption{Example of fit parameters from the initialisation file (Sect.~\ref{sec:init_file}). Only the last column (Value) should be edited.}
\label{tab:fit}
\begin{center}
{
\small
\begin{tabular}{lll}  \hline
Group~~~~Subgroup              & ~~~~Parameter  & \!\!\!\!\!\!\!\!\!\!Multi-entry?~~~Value\\\hline
{\tt UsineFit} @ {\tt Config}    &@ {\tt Minimiser}        &@ {\tt M=0} @ {\tt Minuit2}\\
{\tt UsineFit} @ {\tt Config}    &@ {\tt Algorithm}        &@ {\tt M=0} @ {\tt combined}\\
{\tt UsineFit} @ {\tt Config}    &@ {\tt \dots}            & \\
{\tt UsineFit} @ {\tt TOAData}   &@ {\tt QtiesExpsEType}   &@ {\tt M=0} @ {\tt B/C:AMS:KR}\\
{\tt UsineFit} @ {\tt TOAData}   &@ {\tt ErrType}          &@ {\tt M=0} @ {\tt kERRCOV:\$USINE/inputs/CRDATA\_COVARIANCE/}\\
{\tt UsineFit} @ {\tt TOAData}   &@ {\tt EminData}         &@ {\tt M=0} @ {\tt 1.e-5}\\
{\tt UsineFit} @ {\tt TOAData}   &@ {\tt EmaxData}         &@ {\tt M=0} @ {\tt 1.e10}\\
{\tt UsineFit} @ {\tt TOAData}   &@ {\tt TStartData}       &@ {\tt M=0} @ {\tt 1950-01-01\_00:00:00}\\
{\tt UsineFit} @ {\tt TOAData}   &@ {\tt TStopData}        &@ {\tt M=0} @ {\tt 2100-01-01\_00:00:00}\\
{\tt UsineFit} @ {\tt FreePars}  &@ {\tt CRs}              &@ {\tt M=1} @ {\tt HalfLifeBETA\_10Be:NUISANCE,LIN,[1.3,1.5],1.387,0.012}\\
{\tt UsineFit} @ {\tt FreePars}  &@ {\tt DataErr}          &@ {\tt M=1} @ {\tt SCALE\_AMS02\_201105201605\_\_BC:NUISANCE,LIN,[-10,10],0.,1.}\\
{\tt UsineFit} @ {\tt FreePars}  &@ {\tt DataErr}          &@ {\tt M=1} @ {\tt UNF\_AMS02\_201105201605\_\_BC:NUISANCE,LIN,[-10,10],0.,1.}\\
{\tt UsineFit} @ {\tt FreePars}  &@ {\tt Modulation}       &@ {\tt M=1} @ {\tt phi\_AMS02\_201105201605\_:NUISANCE,LIN,[0.3,,1.1],0.73,0.2}\\
{\tt UsineFit} @ {\tt FreePars}  &@ {\tt SrcSteadyState}   &@ {\tt M=1} @ {\tt alpha:NUISANCE,LIN,[1.7.,2.5],2.3,0.1}\\
{\tt UsineFit} @ {\tt FreePars}  &@ {\tt Transport}        &@ {\tt M=1} @ {\tt eta\_t:FIT,LIN,[-3,3],0.,0.1}\\
{\tt UsineFit} @ {\tt FreePars}  &@ {\tt Transport}        &@ {\tt M=1} @ {\tt delta:FIT,LIN,[0.2,0.9],0.6,0.02}\\
{\tt UsineFit} @ {\tt FreePars}  &@ {\tt XSection}         &@ {\tt M=1} @ {\tt EnhancePowHE\_ALL:NUISANCE,LIN,[0.,2.],1.,1.}\\
{\tt UsineFit} @ {\tt FreePars}  &@ {\tt XSection}         &@ {\tt M=1} @ {\tt Norm\_16O+H:NUISANCE,LIN,[0.7,1.3],1.,0.05}\\
{\tt UsineFit} @ {\tt FreePars}  &@ {\tt XSection}         &@ {\tt M=1} @ {\tt Norm\_12C+H->11B:NUISANCE,LIN,[0.5,2.],1.,0.2}\\
{\tt UsineFit} @ {\tt FreePars}  &@ {\tt XSection}         &@ {\tt M=1} @ {\tt EAxisScale\_12C+H->11B:NUISANCE,LIN,[0.5,1.5],1.,0.5}\\
\hline\hline
\end{tabular}
}
\end{center}
\vspace{-.4cm}
\end{table*}

The implementation of minimisers in \usine{} is based on the {\tt ROOT::Math::Minimizer} wrapper. A minimisation run, based on $\chi^2$ (see Sect.~\ref{sec:code}), requires additional parameters, as illustrated in Table~\ref{tab:fit}. The syntax of these parameters is detailed in the documentation, as it can be quite involved for free parameters.

\begin{itemize}

   \item {\tt UsineFit@Config}: selection of the minimiser configuration and its parameters. I use as default {\tt minuit2} (see online manual\footnote{\url{https://root.cern.ch/guides/minuit2-manual}}) and {\sc minos} can be switched on to properly get uncertainties on the free parameters.

   \item {\tt UsineFit@TOAData}: selection of CR data on which to fit, and whether to use covariance or not (if exists, see \ref{sec:crdata}). For instance, in Table~\ref{tab:fit}, I fit on {\tt B/C:AMS:KR}, that is B/C$(R)$ from AMS data using a covariance matrix to be searched in the directory {\tt \$USINE/inputs/CRDATA COVARIANCE}, with no sub-selection on the energy and time ranges.

   \item {\tt UsineFit@FreePars}: geometry, ISM, source, and transport parameters can all in principle be set as free parameters. In addition, modulation parameters and data error uncertainties (for all data periods used in the fit) can be added as nuisance parameters\footnote{In both cases, the name of the parameter is built internally from the experiment name (removing special characters). Run {\tt ./bin/usine\_run -m1} on your initialisation file to list what parameters can be left free for your configuration.}. The syntax of a fit or nuisance parameter is {\tt name:X,Y,[min,max],init,sigma}. Use {\tt X = FIT} to declare as a fit parameter or {\tt X = NUISANCE} to assume a Gaussian distributed parameter ${\cal N}(\mu) =$ {\tt init}, $\sigma =$ {\tt sigma}, see Eq.~(\ref{eq:chi2_nuis}); use {\tt Y = LIN {\rm or} LOG} to sample the parameter as {\tt name} or $\log_{10}$({\tt name}). Note that values outside {\tt [min,max]} are strongly penalised. In Table~\ref{tab:fit}, I have for instance:
   \begin{itemize}
      \item fit parameters: the universal slope {\tt alpha} of the source spectrum (parameter {\tt SrcSteadyState}), and the transport coefficients {\tt eta\_t} and {\tt delta} (parameters {\tt Transport});
      \item nuisance parameters: $t_{1/2}^{^{10}\rm Be}$ (parameter {\tt CRs}), solar modulation level of the specified AMS data (parameter {\tt Modulation}), inelastic and production cross-sections (parameters {\tt XSection}), specific error systematics (parameters {\tt DataErr})\footnote{Any systematic uncertainties in the covariance matrix declared as a nuisance parameter (here {\tt SCALE} and {\tt UNF}) contributes to the $\chi^2$ as a nuisance (\ref{eq:chi2_nuis}) and is no longer dealt with in the covariance (\ref{eq:cov}).}.
  \end{itemize}
\end{itemize}


\section{Executable, options, and outputs: run examples \label{sec:run_examples}}

The executable is a command-line with various options and arguments,

   {\tt ./bin/usine -option arg1 arg2...}

\noindent where {\tt option} redirects to specific runs (standard calculation, extra plots, minimisation), {\tt arg1} is always the initialisation file (see Sect.~\ref{sec:init_file}), and {\tt arg2} is generally an output directory.

\paragraph{Options}
The five families of options are:
\begin{itemize}
   \item {\tt ./bin/usine -l}: local flux calculation showing only selected quantities;
   \item {\tt ./bin/usine -m}: minimisation-related options, with {\tt -m1} to list fit-able parameters, and {\tt -m2} to perform minimisations;
   \item {\tt ./bin/usine -e}: extra calculations and comparison plots via an interactive session;
   \item {\tt ./bin/usine -i}: input files/properties to display CR data, cross sections, etc.;
   \item {\tt ./bin/usine -t}: test functions for \usine{}.
\end{itemize}

Each option comes with a mandatory list of arguments, which is self-explanatory for all the options above: the {\tt USAGE} line gives the arguments, and the line {\tt EXAMPLE} gives default values for all arguments (that can be run directly).

\paragraph{Initialisation files}
Several initialisation files are shipped with this release: {\tt inputs/init.TEST.par} is a test-only file, used to check the installation of \usine{} classes and all models. We also provide one file per \usine{} model: {\tt inputs/init.Model0D.par}, {\tt inputs/init.Model1D.par}, and {\tt inputs/init.Model2D.par} for the LB, 1D, and 2D models respectively. Their settings allow to roughly match B/C AMS-02 data, and they can be used as a starting point for more elaborate analyses within these models. Note that with the choice of parameters, the 1D and 2D models differ by less than $0.1\%$, and the 1D and LB models by less than 2\% over the AMS-02 B/C rigidity range. We also provide in the documentation additional configurations resulting for the full B/C analysis performed in \cite{2019arXiv190408917G}.

\paragraph{Outputs}
The outputs of any given run are \ascii{} files and images saved in the user-selected output directory (e.g., {\tt \$USINE/output/} in the examples below). Available \ascii{} files are a log file of the run ({\tt usine.last\_run.log}), a backup of the initialisation file last run ({\tt usine.last\_run.init.par}), and many result files for propagated fluxes and ratios ({\tt local\_fluxes*.out}). Most \usine{} options also produce \ascii{} files (description and timestamp in the header) associated to the shown plots (saved as {\tt .pdf} and {\tt .png}, and \rootcern{} macros {\tt .C} files). Outputs related to the calculation of IS and TOA fluxes are similar for all models, only differing in their name. The spatial distribution of fluxes can be calculated for 1D and 2D models, but it is not an output yet in this release.

\subsection{Local fluxes: {\tt ./bin/usine -l}\label{sec:option_l}}

\begin{figure*}[!th]
\centering
\includegraphics[width=0.49\textwidth]{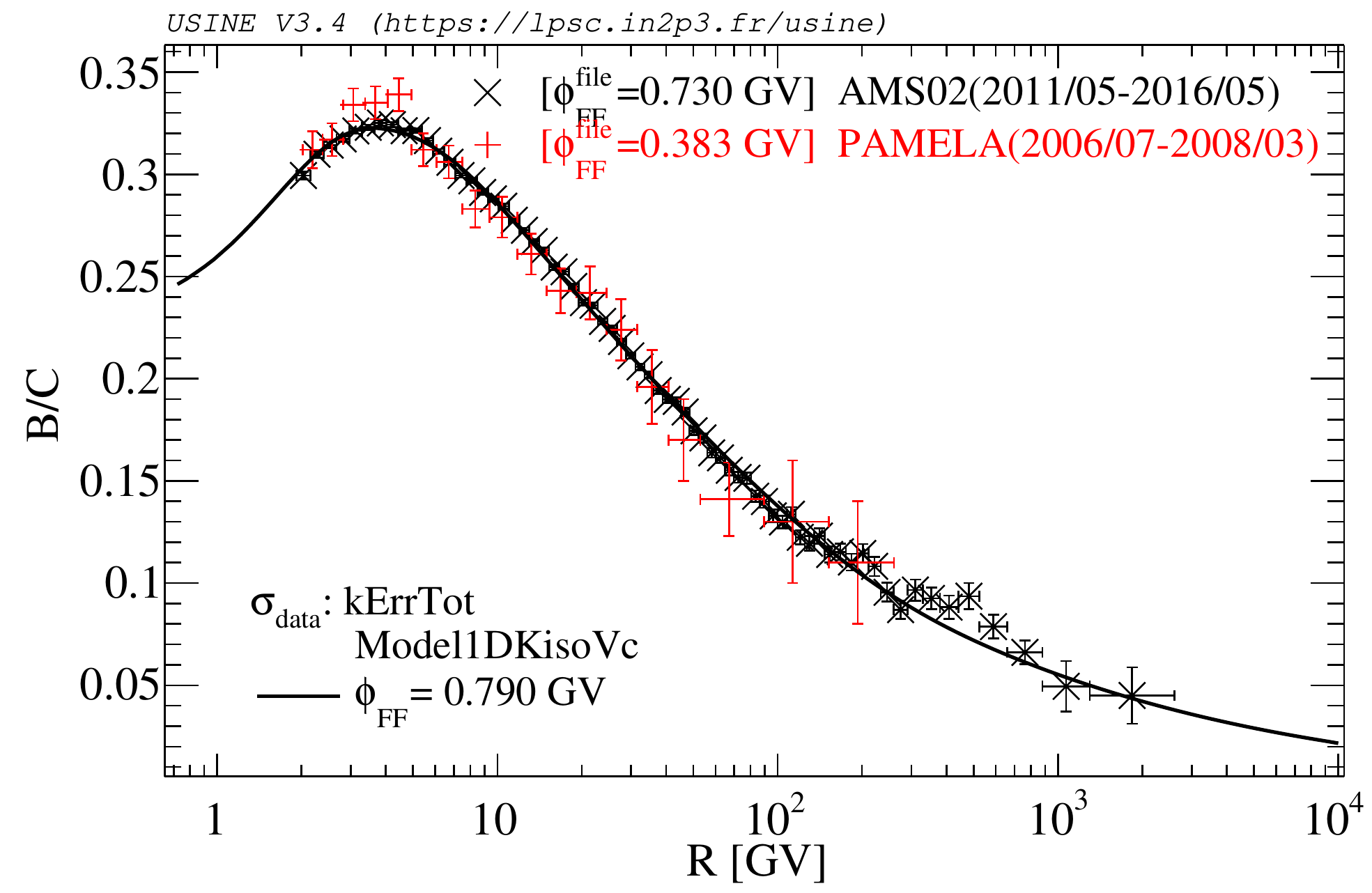}
\includegraphics[width=0.49\textwidth]{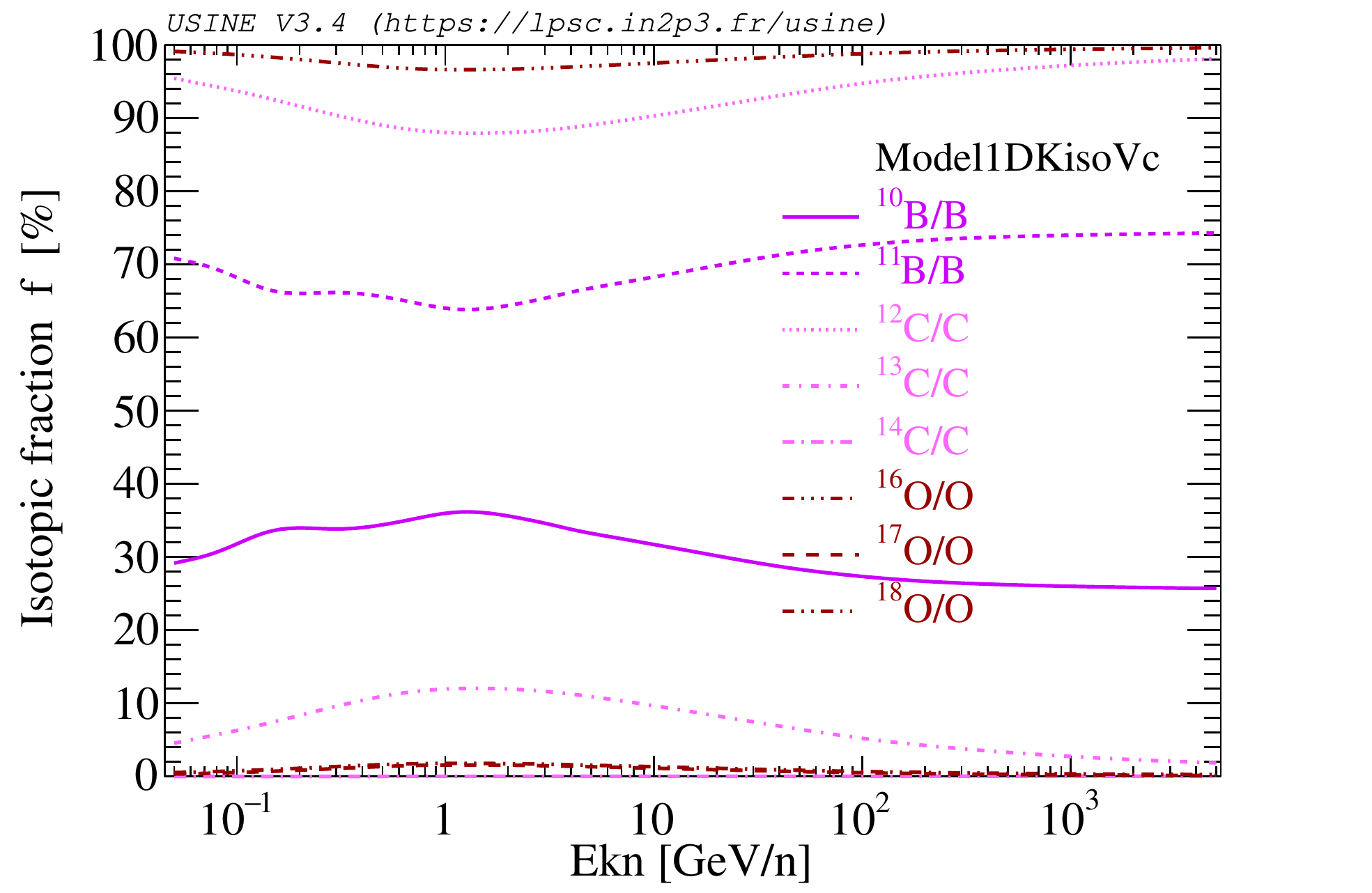}
\includegraphics[width=0.49\textwidth]{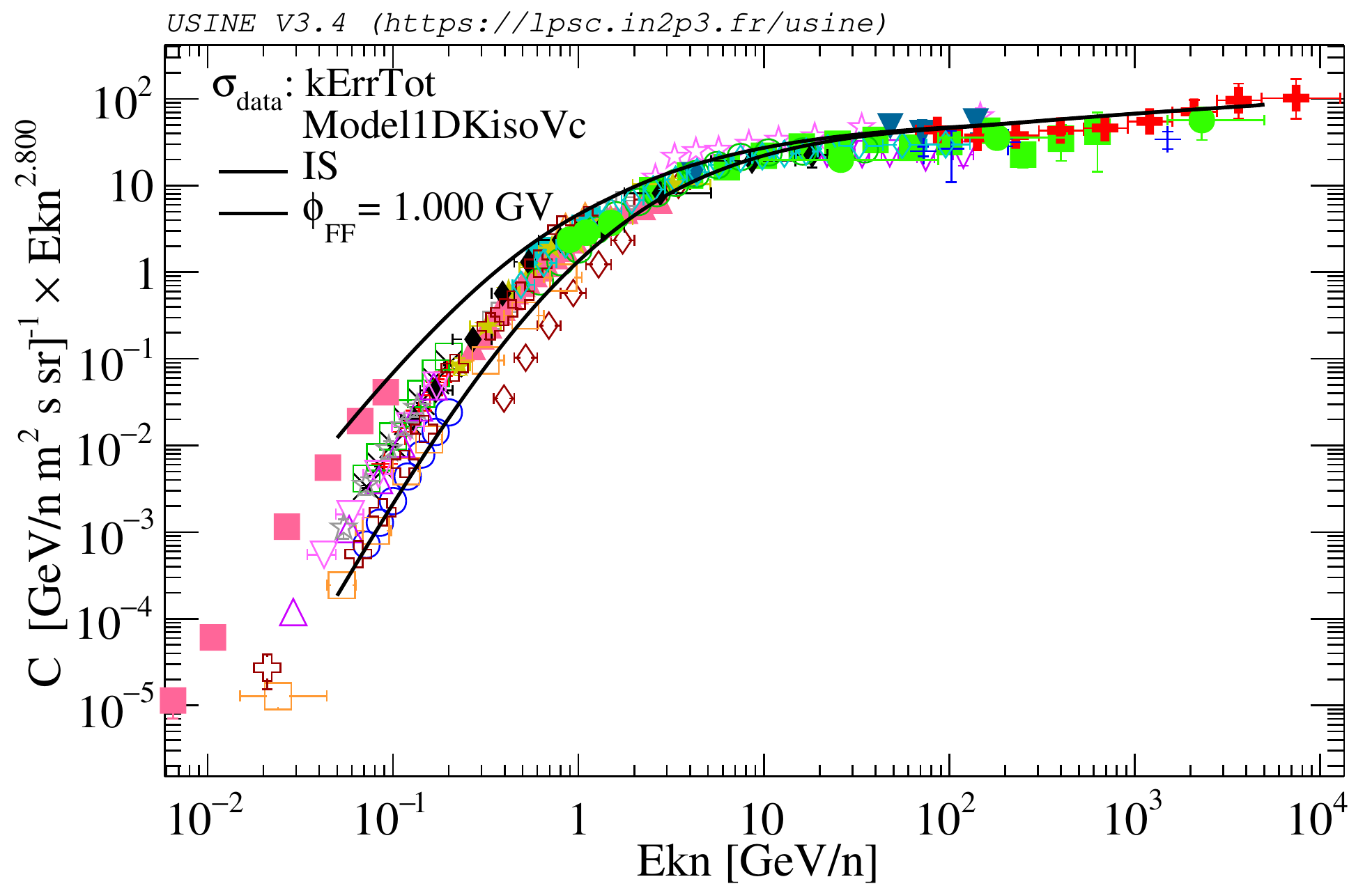}
\includegraphics[width=0.49\textwidth]{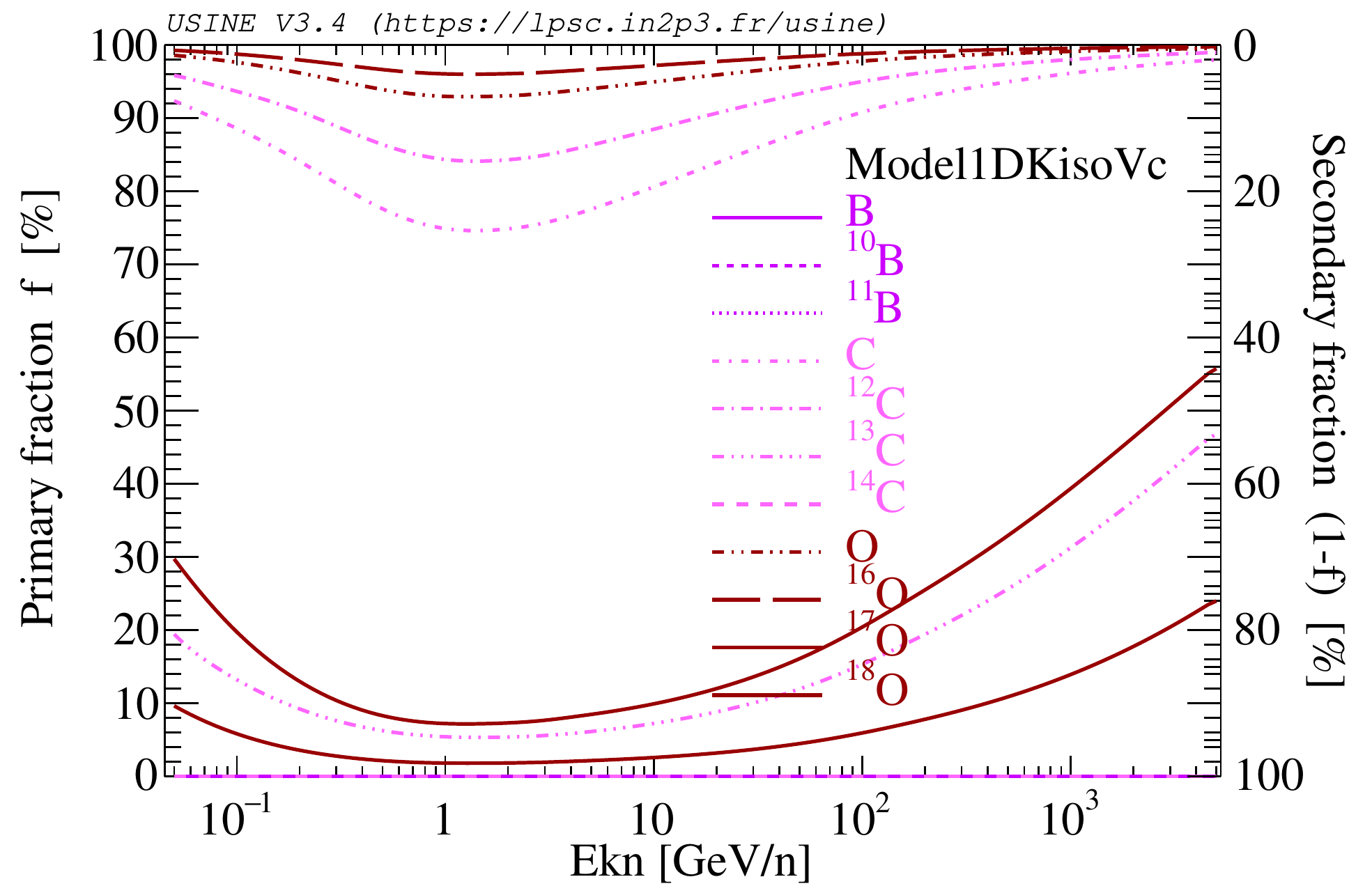}
\caption{Plots from option {\tt -l}. The left panels illustrate the  display of a ratio and a flux in different energy unit ($R$ and $E_{k/n}$). For all user-selected CRs, \usine{} also displays the energy-dependent fraction of all associated isotopes (top right), and the fraction of primary and secondary origin for each corresponding isotopes and elements (bottom right). See corresponding command line and full figure description in Sect.~\ref{sec:option_l}.}
\label{fig:option_l}
\end{figure*}

This option runs and calculates local fluxes, ratios, etc. for several modulation levels. Many plots and files are produced from the command line:\\
{\tt \small ./bin/usine -l inputs/init.Model1D.par \$USINE/output "B/C,B,C,O:KEKN:0.,1.;B/C:kR:0.79" 2.8 1 1 1}\\

Figure~\ref{fig:option_l} shows, from top to bottom, (i) B/C vs $R$ at selected modulation level along with selected CR data, (ii) C multiplied by $E_{k/n}^{2.8}$ vs $E_{k/n}$, (iii) isotopic fractions (form IS fluxes) vs $E_k{/n}$ for all isotopes involved in the selection, and (iv) primary (left axis) or secondary ($100-$primary, right axis) content of IS fluxes vs $E_{k/n}$ for all quantities involved in the selection.

\subsection{Extra plots: {\tt ./bin/usine -e}\label{sec:option_e}}
\begin{figure*}[!th]
\centering
\includegraphics[width=0.49\textwidth]{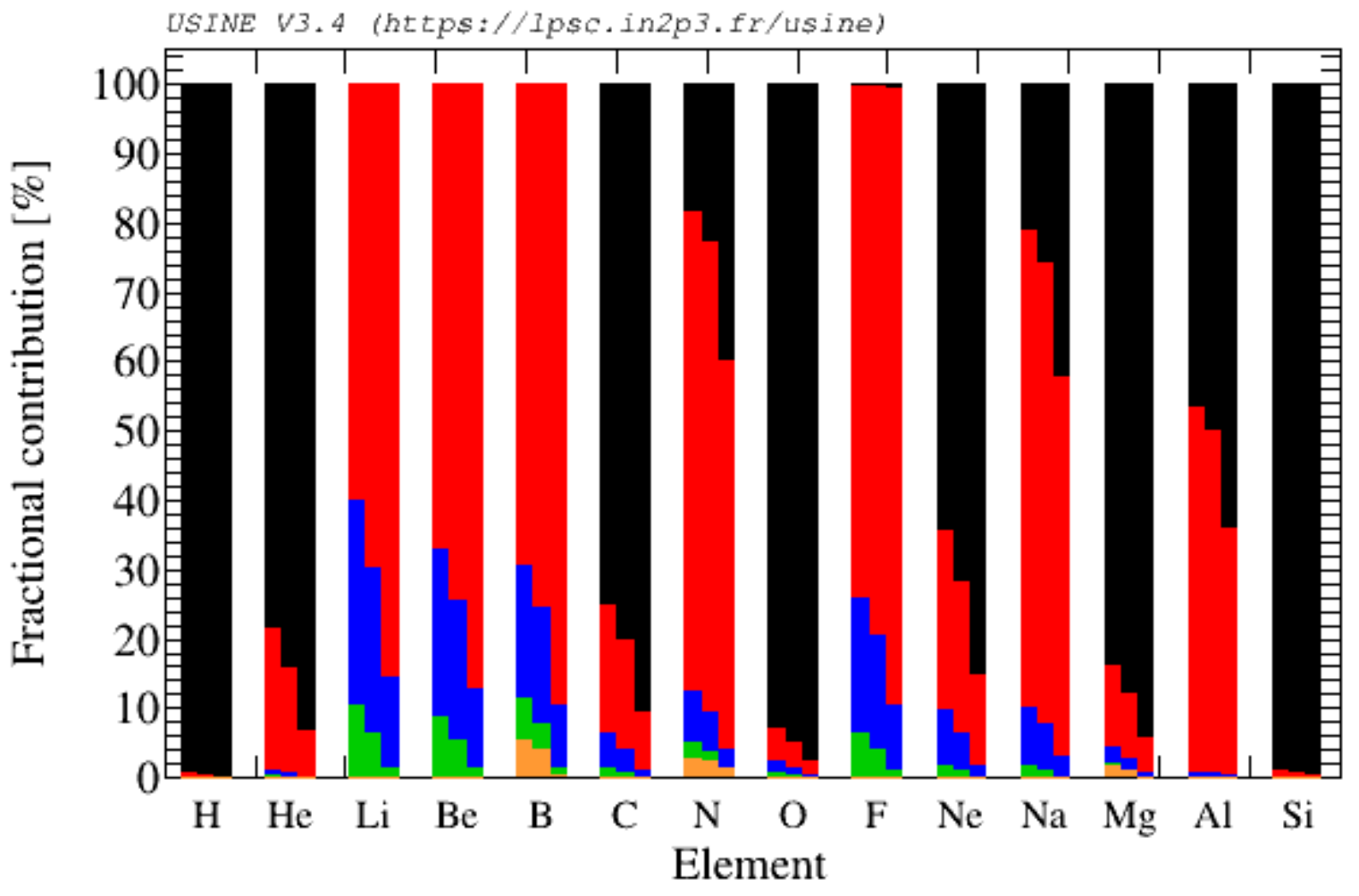}
\includegraphics[width=0.49\textwidth]{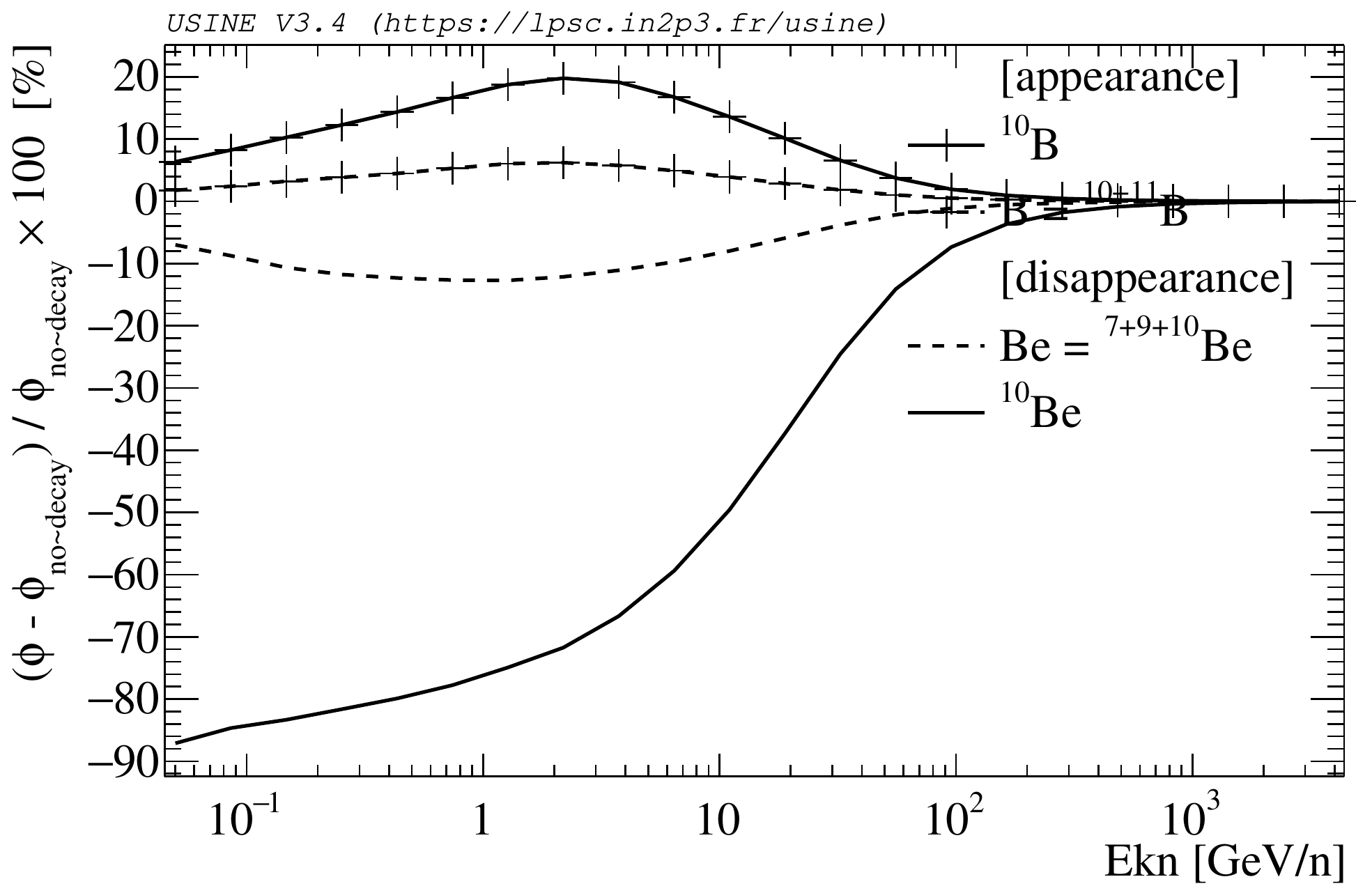}
\includegraphics[width=0.49\textwidth]{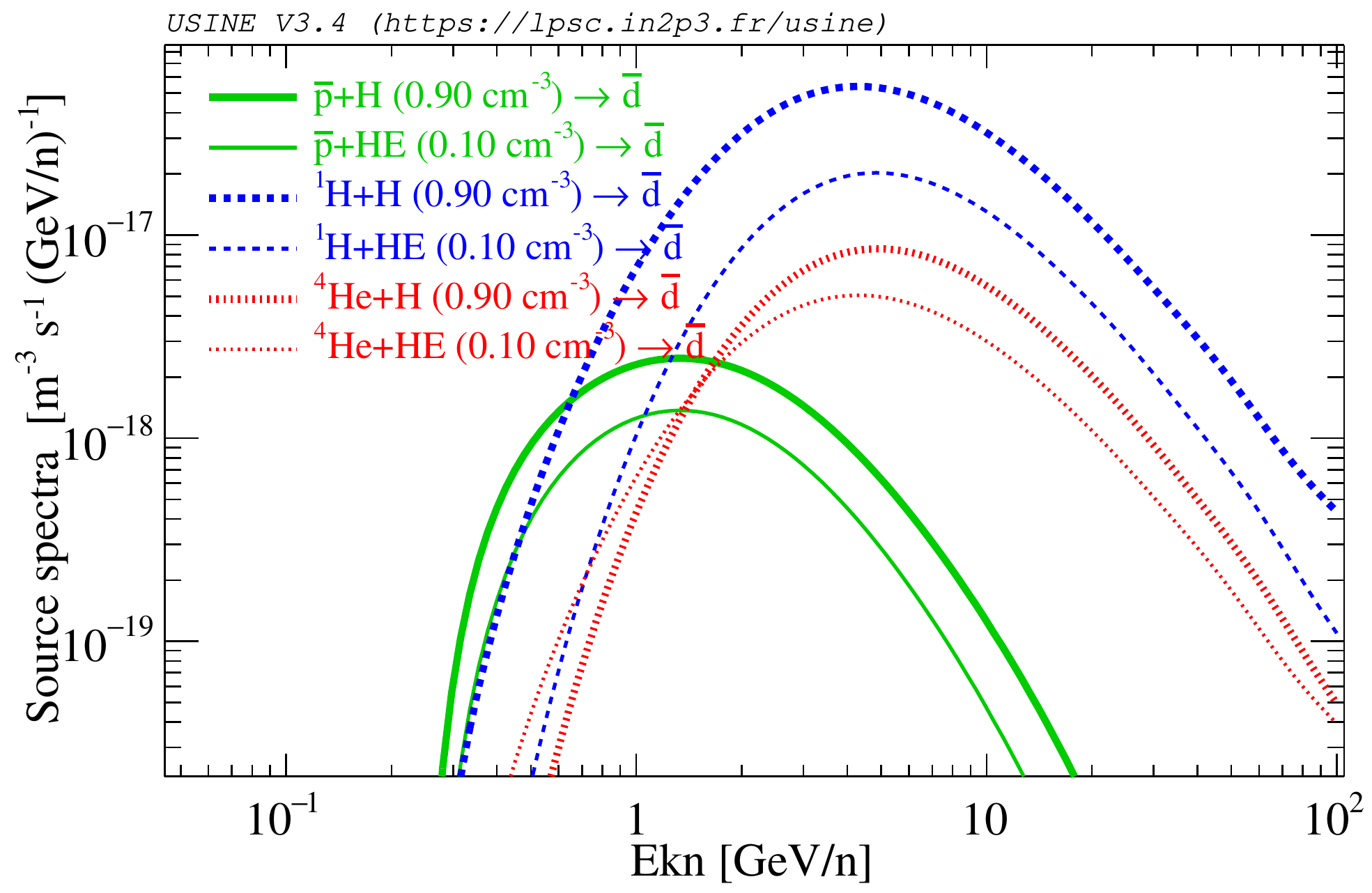}
\includegraphics[width=0.49\textwidth]{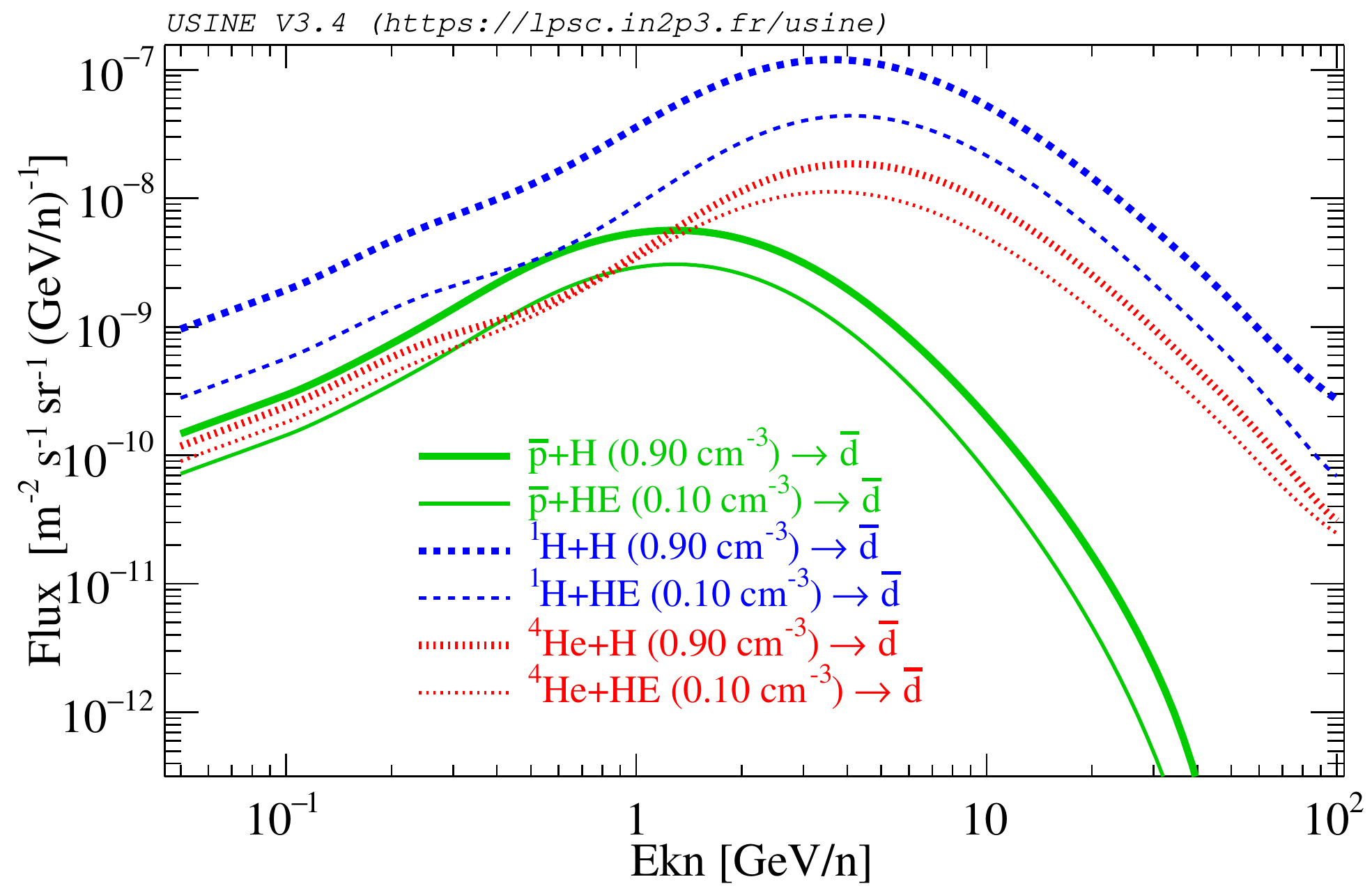}
\includegraphics[width=0.49\textwidth]{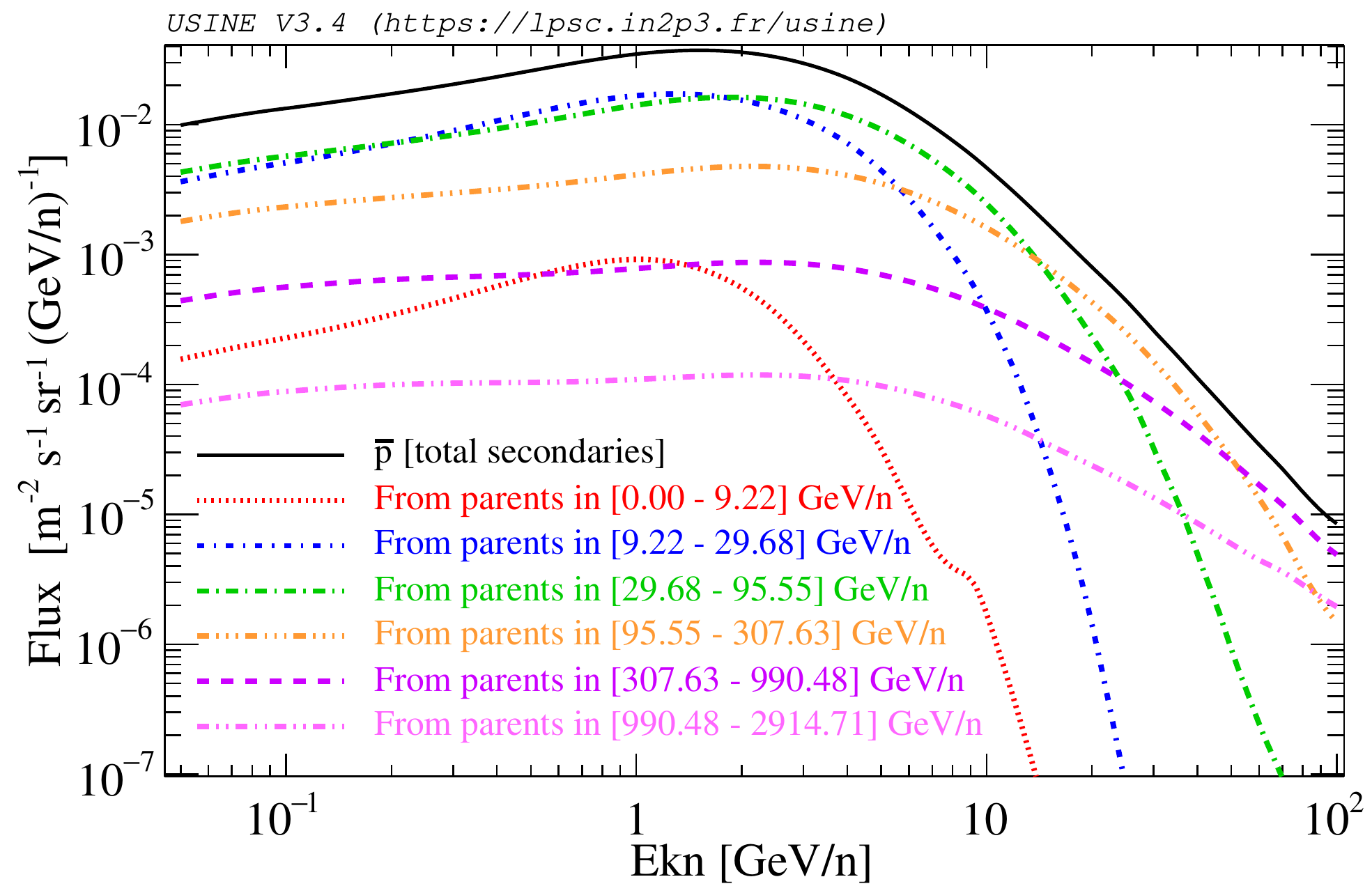}
\includegraphics[width=0.49\textwidth]{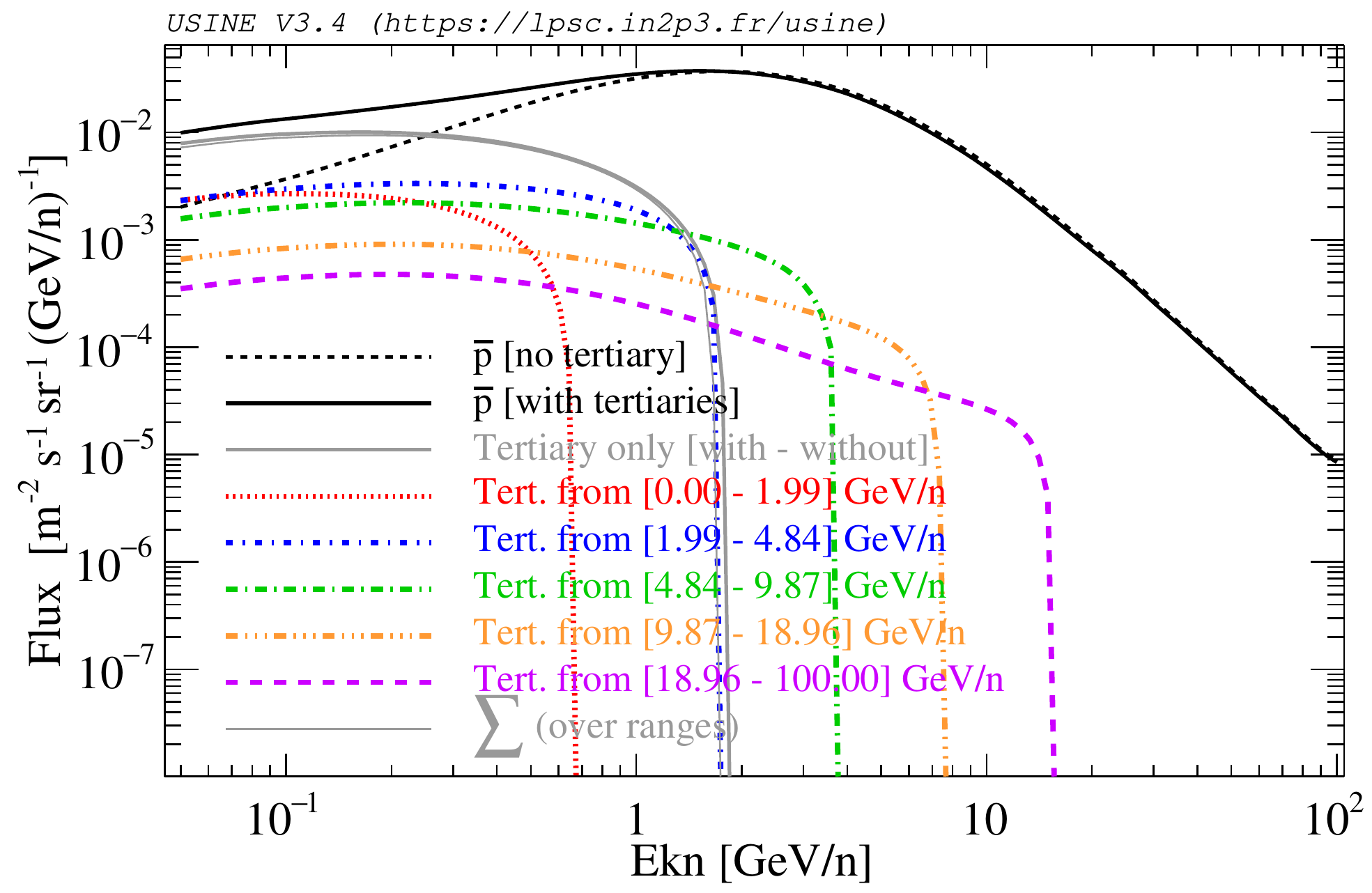}
\caption{Sample of plots from option {\tt -e} and sub-options {\tt D}. {\bf Top panels:} elemental fractional origins at three energies (left panel, see text for details), and impact of decay on propagated CR flux for an unstable species, its daughter, and associated elements (right panel). {\bf Middle panels:} anti-deuteron contributions from CR projectiles and ISM targets separately at source (left panel, primary species folded to cross sections), or after propagation (right panel). {\bf Bottom panels:} anti-proton spectrum broken-down in contributions from various energy ranges for parents (left-panel) or tertiaries (right panel). See corresponding command line and full figure description in Sect.~\ref{sec:option_e}.}
\label{fig:option_eD}
\end{figure*}
\begin{figure*}[!th]
\centering
\includegraphics[width=0.49\textwidth]{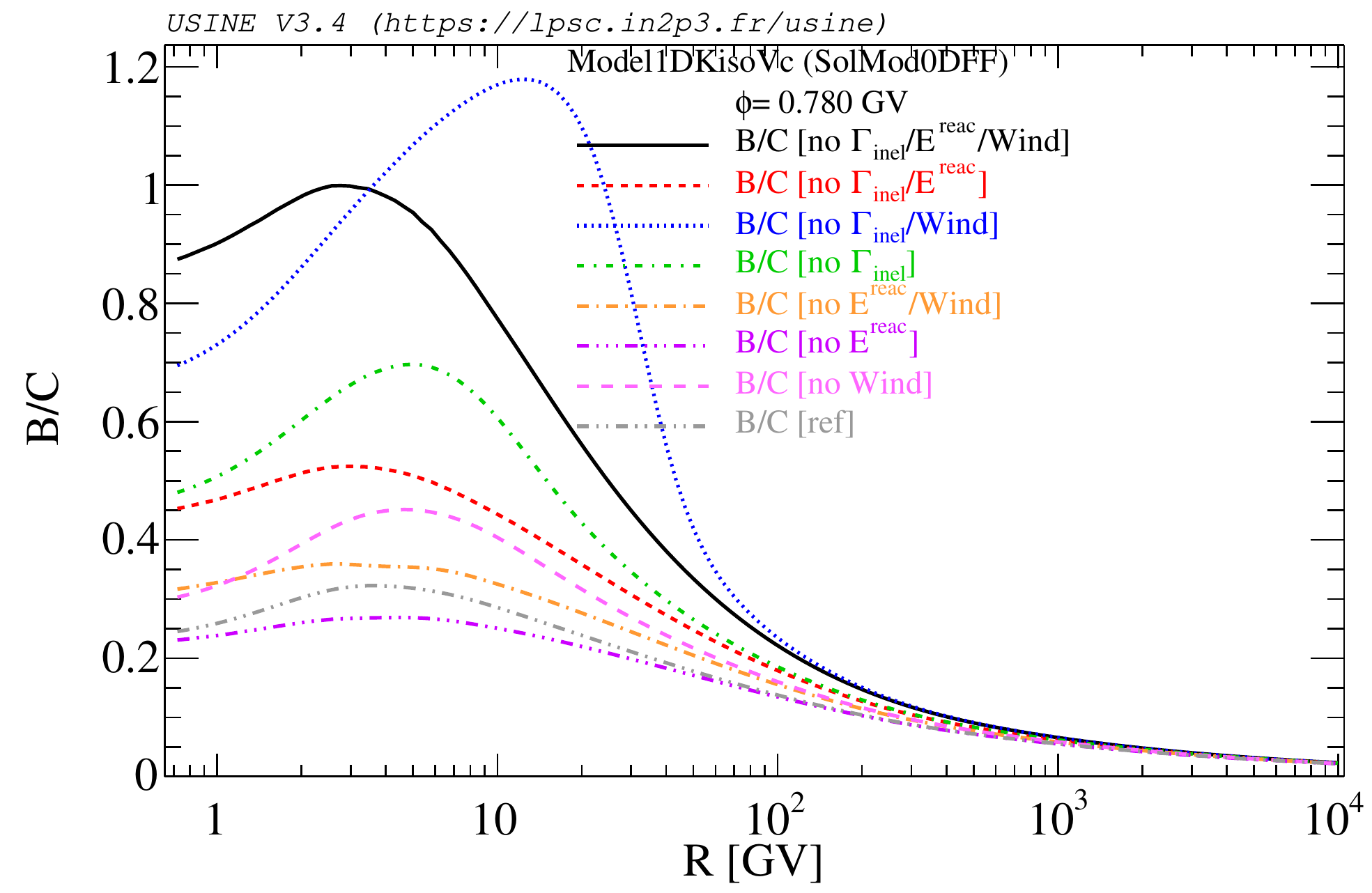}
\includegraphics[width=0.49\textwidth]{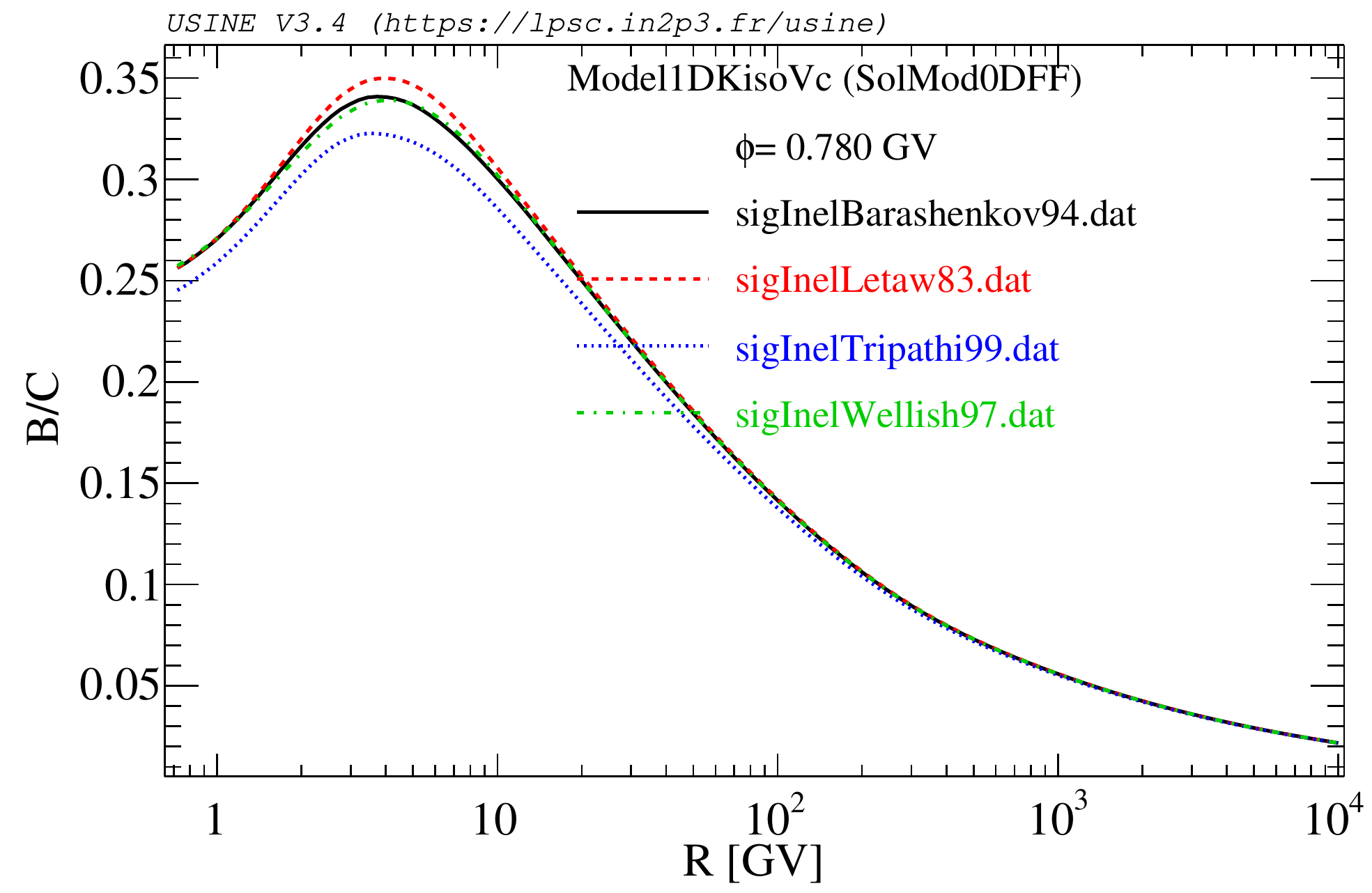}
\includegraphics[width=0.49\textwidth]{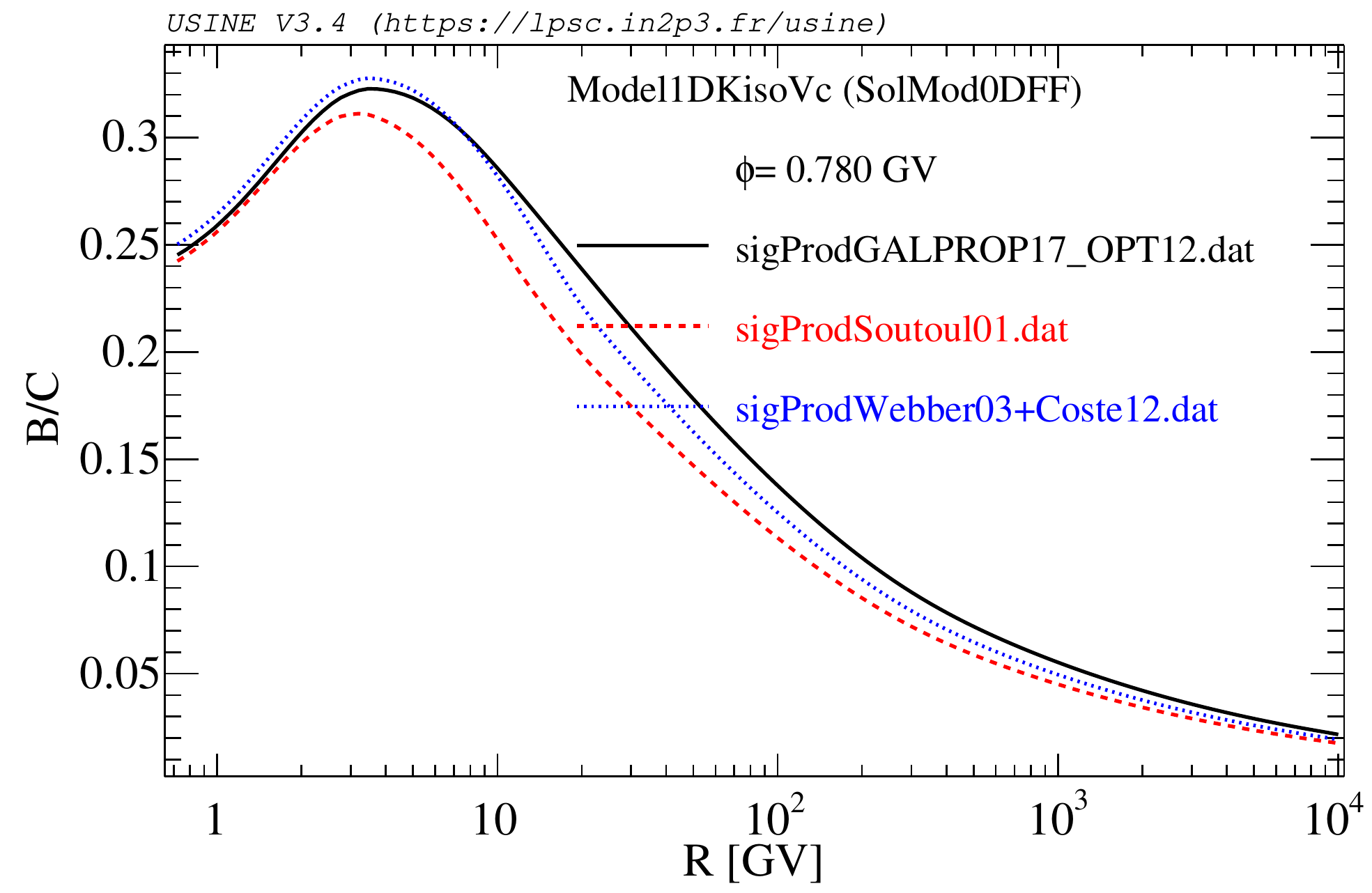}
\includegraphics[width=0.49\textwidth]{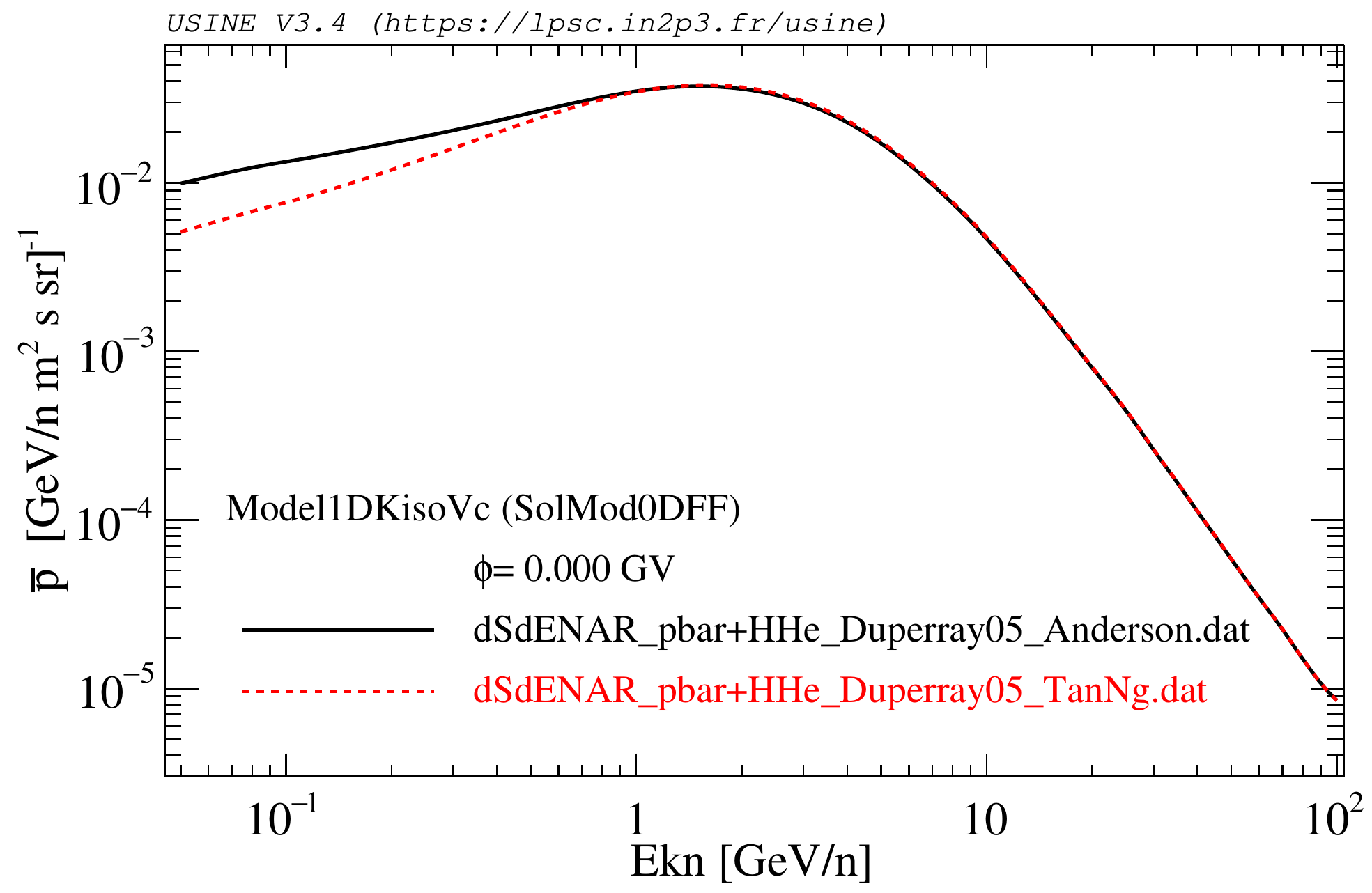}
\caption{Plots from option {\tt -e} and sub-options {\tt E}. Calculations switching on and off several propagation effects for B/C (top left); effect of using different sets of inelastic (top right) and production (bottom left) cross sections on B/C; effects of using two different schemes for non-annihilating rescattering cross sections for anti-protons (bottom right). See corresponding command line and full figure description in Sect.~\ref{sec:option_e}.}
\label{fig:option_eE}
\end{figure*}

An interactive session in which the user can loop on sub-options\footnote{After display pops-ups, just quit \rootcern{} in any window to go back to the interactive session.} is started via the command line:\\
{\tt \small ./bin/usine -e   inputs/init.Model1D.par \$USINE/output\,1\,1\,1}

\paragraph{Plots from sub-options {\tt -D}}
Figure~\ref{fig:option_eD} shows from top left to bottom right:
\begin{itemize}
   \item {\tt D0}: relative contributions per production process for elemental fluxes (isotopes not shown) at 1, 10, and 100 GeV/n: primary (black), secondary (1, 2, and $>2$ steps in red, blue, and green), radioactive (orange);
   \item {\tt D3}: relative fraction of IS flux in excess or disappeared for $\beta$-radioactive isotope $^{10}$Be and its daughter $^{10}$B. The relative fraction for the element Be and B are also shown: the impact of decay is diluted by the other stable isotopes of the element);
   \item {\tt D5}: separated source terms (before propagation) for CR+ISM reactions leading to $\bar{d}$;
   \item {\tt D6}: separated $\bar{d}$ contributions (after propagation) for CR+ISM reactions;
   \item {\tt D7}: secondary $\bar{p}$ contributions from different primary CR energy ranges;
   \item {\tt D8}: tertiary $\bar{p}$ contribution. The black lines show the calculated flux without (dotted) and with (solid) the tertiary contribution: non-annihilating rescattering of $\bar{p}$ on the ISM deplete $\bar{p}$ at high energies and redistribute them at lower energies. The various colours show the contributions of various energy ranges.
\end{itemize}

\paragraph{Plots from sub-options {\tt -E}}
Figure~\ref{fig:option_eE} shows from top left to bottom right:
 From top left to bottom right: 
\begin{itemize}
   \item {\tt E2}: comparison plot switching on/off physics effects during propagation (all other parameters being equal). In this example for B/C(R), the inelastic cross sections, reacceleration, and galactic wind have been switched-off in turns or together, compared to the reference (grey line).;
   \item {\tt E3}: comparison plot using different inelastic cross-section parametrisations for nuclei;
   \item {\tt E4}: comparison plot using different production cross-section parametrisations for nuclei;
   \item {\tt E8}: comparison plot using different tertiary differential (redistribution) cross sections for $\bar{p}$.
\end{itemize}

\subsection{Extra plots: {\tt ./bin/usine -m}\label{sec:option_m}}

\begin{figure}[!th]
\centering
\includegraphics[width=\columnwidth]{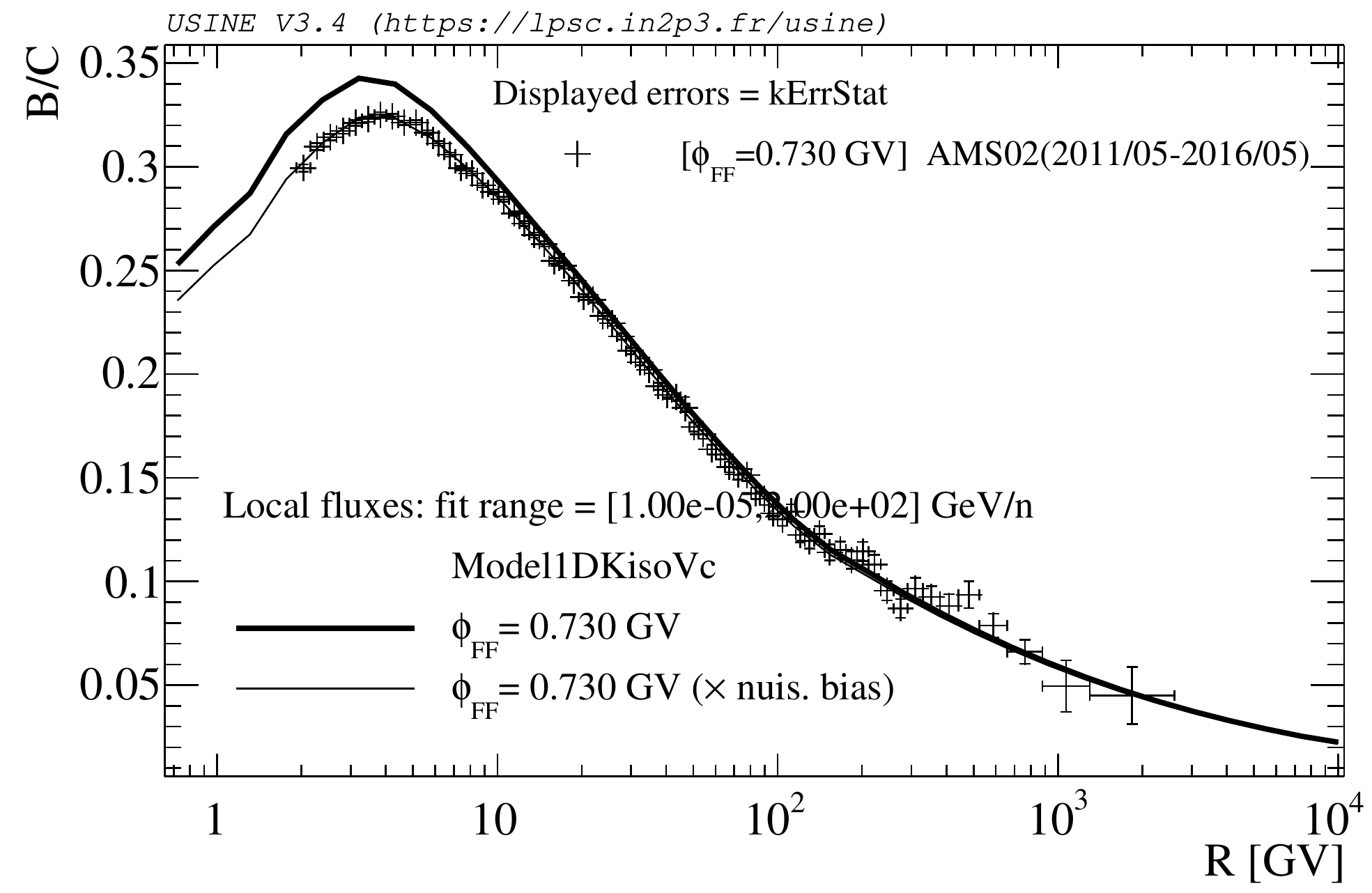}
\includegraphics[width=\columnwidth]{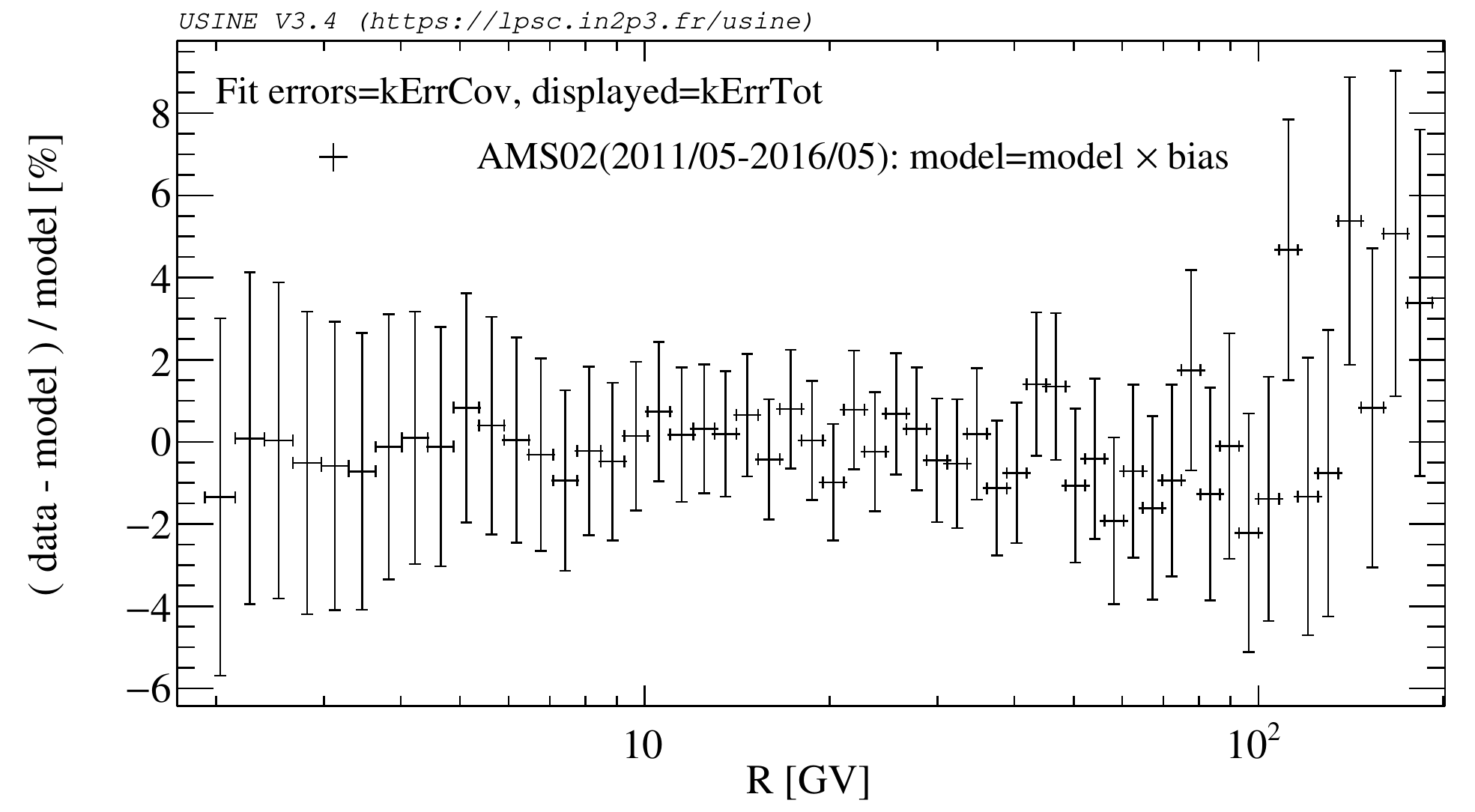}
\includegraphics[width=\columnwidth]{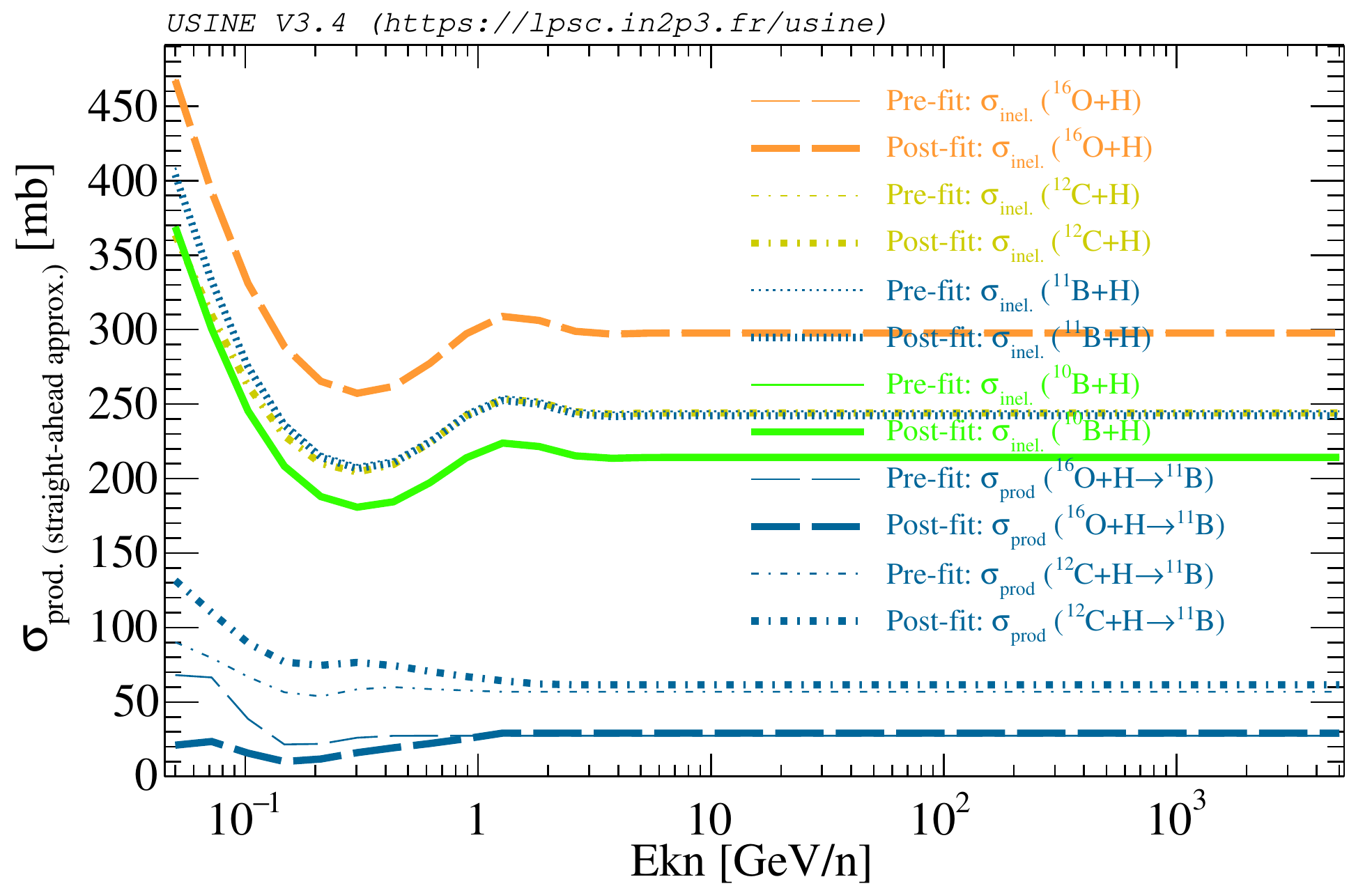}
\caption{Plots from minimisation option {\tt -m2}. In this example, based on {\tt inputs/init.TEST.par} with modifications, the top panel illustrates the presence of a nuisance parameter on the data systematics (thin vs thick line), the middle panel shows residuals between data and best-fit model, and the bottom panel shows a comparison of pre- and post-fit cross-section nuisance parameters. See corresponding command line and full figure description in Sect.~\ref{sec:option_m}.}
\label{fig:option_m}
\end{figure}

Minimisation runs with fit parameters, and possibly nuisance parameters and covariance matrix are performed with the command line:\\
{\tt \small ./bin/usine -m2  inputs/init.TEST.par  \$USINE/output\_fit   1    1   1}

In addition to the information relative to the fit, several plots are displayed on screen, as illustrated in Fig.~\ref{fig:option_m}, from top to bottom:
\begin{itemize}
   \item Best-fit obtained along with the data selected for display. If nuisance parameters are used for systematics errors of the fit data, both the model (thick line) and the biased model (thin line) are shown, with ${\rm bias}(R_k)=\Pi_{l=0…n}(1+\nu_l \times \sigma_l(R_k))$.
   \item Residual between (biased)-model and data selected for the fit. The error bars correspond to the relative data errors.
   \item If specific cross sections are enabled as nuisance parameters (see Sect.~\ref{sec:fit}), show the pre-fit (thin lines/small symbols) and post-fit cross sections (thick lines/large symbols) for comparison.
\end{itemize}


\section{Conclusions \label{sec:conclusion} }

The {\sc usine} code is a library for galactic cosmic-ray propagation, solving the diffusion equation for several semi-analytical models (LB, 1D-, and 2D-diffusion models). The code structure easily allows to add more complicated models, because many inputs, functions (to calculate primaries, secondaries, tertiaries\dots), and displays are of generic design and do not belong to any particular model.

A run configuration (ingredients, fixed and free parameters) is fully specified by a single \ascii{} file. To help navigate among the various inputs and keywords, an online documentation details all the files content and format, as well as the many keywords usage. Almost any model parameter (geometry, source, diffusion, ISM, Solar modulation level) can be set as a free parameter for minimisation studies. In addition, nuisance parameters on CR properties (e.g. decay time), cross sections (normalisation, high-energy dependence, etc.), and CR data (or correlation matrix) are enabled. These specificities were developed and taken advantage of to provide sound statistical analyses of recent high-precision AMS-02 data in \cite{2019arXiv190408210D,2019arXiv190408917G,2019arXiv190607119B}.

Thanks to the speed of semi-analytical models, \usine{} can also be used in an interactive mode, in which the calculation of IS fluxes is done once: a simple text interface allows users to loop on displayed fluxes and ratios at any modulation level. Extra plots for isotopic and primary fractions in CR fluxes, impact of decay, comparison of results using several cross-section files, etc. are also provided, and all associated data are saved in \ascii{} files. We stress that fluxes and ratios can be calculated for several energy units (rigidity, kinetic energy per nucleon, etc.), naturally matching units provided in different CR measurements. The latter can be extracted from CRDB \cite{2014A&A...569A..32M}, whose format is directly the one required by \usine{}.

Many improvements are possible for a future release. My wish list would be the inclusion of electrons and positrons in 1D and 2D models following \cite{2017A&A...605A..17B}, the inclusion of dark matter contributions in 2D models (as was present in the second unreleased version of the code), the interface with MCMC engines, add splines as alternative to formulae in calculations, the proper inclusion of EC-decay CRs in propagation, 1D spherically symmetric Solar modulation model, and the extension of CR list and cross-section files for very heavy CRs ($Z>30$).

\section*{Acknowledgements}
I am indebted to my colleagues for their encouragements and contributions on previous unreleased \usine{} versions: F.~Barao, L.~Derome, F.~Donato, A.~Putze, P.~Salati, and R.~Taillet. Thanks a lot to the CRAC team for their feedback and suggestions on this release: G.~B\'elanger, M.~Boudaud, S.~Caroff, Y.~G\'enolini, J.~Lavalle, V.~Poireau, V.~Poulin, S.~Rosier-Lees, P.~I.~Silva Batista, P.~Serpico, and M.~Vecchi. Special thanks to Y. G\'enolini and M. Boudaud for helping find out many bugs, for providing the initialisation files obtained in \cite{2019arXiv190408917G}, and for the new anti-proton cross sections used in \cite{2019arXiv190607119B}. I thank the three anonymous referees for their comments and questions that helped to provide a more complete description of the code. Many thanks to C.~Combet and M.~H\"utten, from the {\sc clumpy} team (\url{https://lpsc.in2p3.fr/clumpy}), for sharing ideas about how to set nicely \usine{} online. Lastly, I am grateful for the IT support of F.~Melot at LPSC. This work has been supported by the ``Investissements d'avenir, Labex ENIGMASS".

\appendix
\section{Main features of \usine{} v3.5 \label{app:v3.5}}

\usine{} v3.5 was developed for the AMS-02 B/C and \pbar{} analyses carried out in \cite{2019arXiv190408210D,2019arXiv190408917G,2019arXiv190607119B}. With respect to v3.4, it has more flexibility in the fit parameters and more plots for minimisation analyses. We list below the main improvements, and refer the reader to the version release notes\footnote{\url{https://dmaurin.gitlab.io/USINE/general_release3.5.html}} for the full changes.

\begin{figure}[!th]
\centering
\includegraphics[width=\columnwidth]{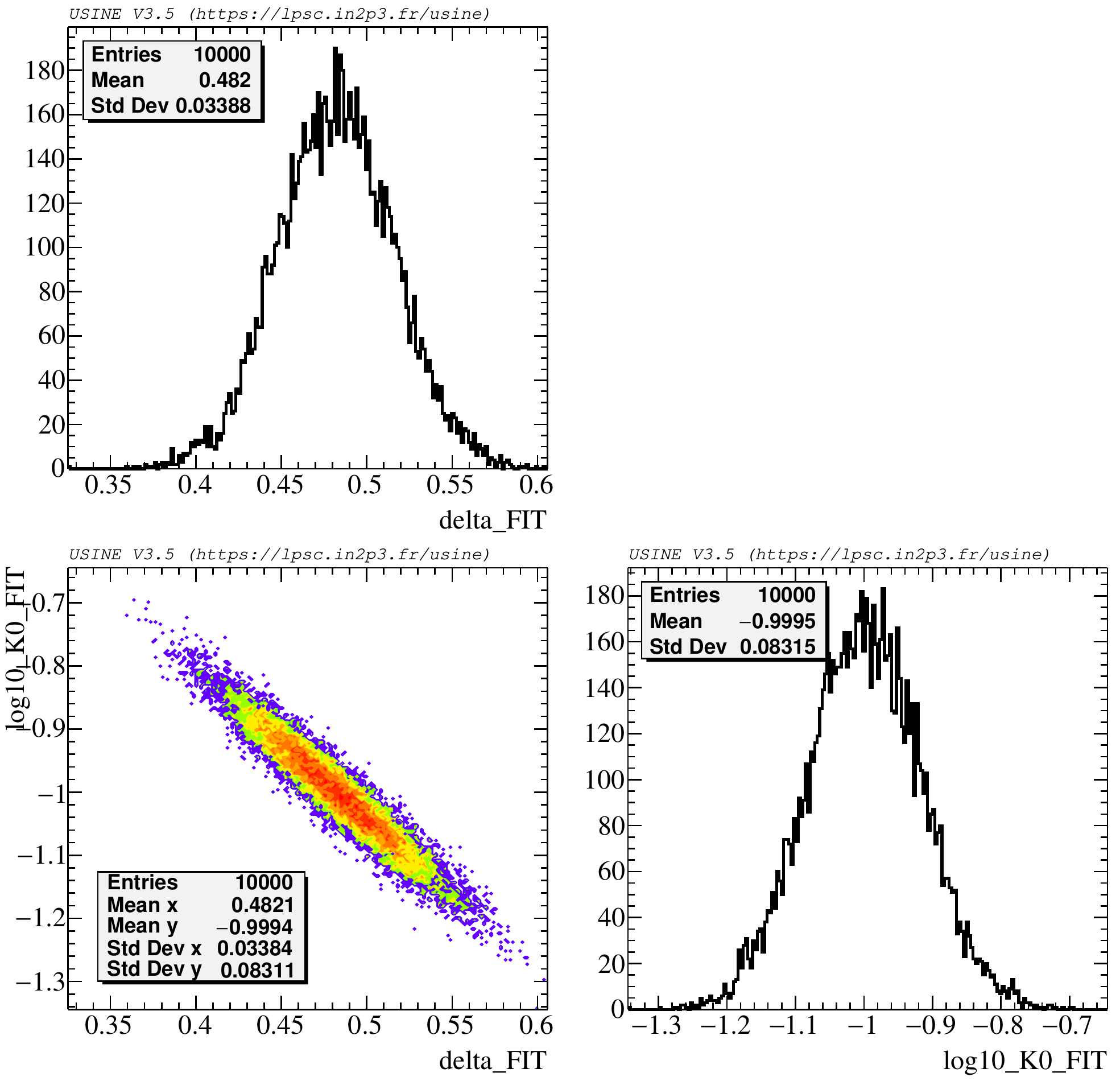}
\includegraphics[width=\columnwidth]{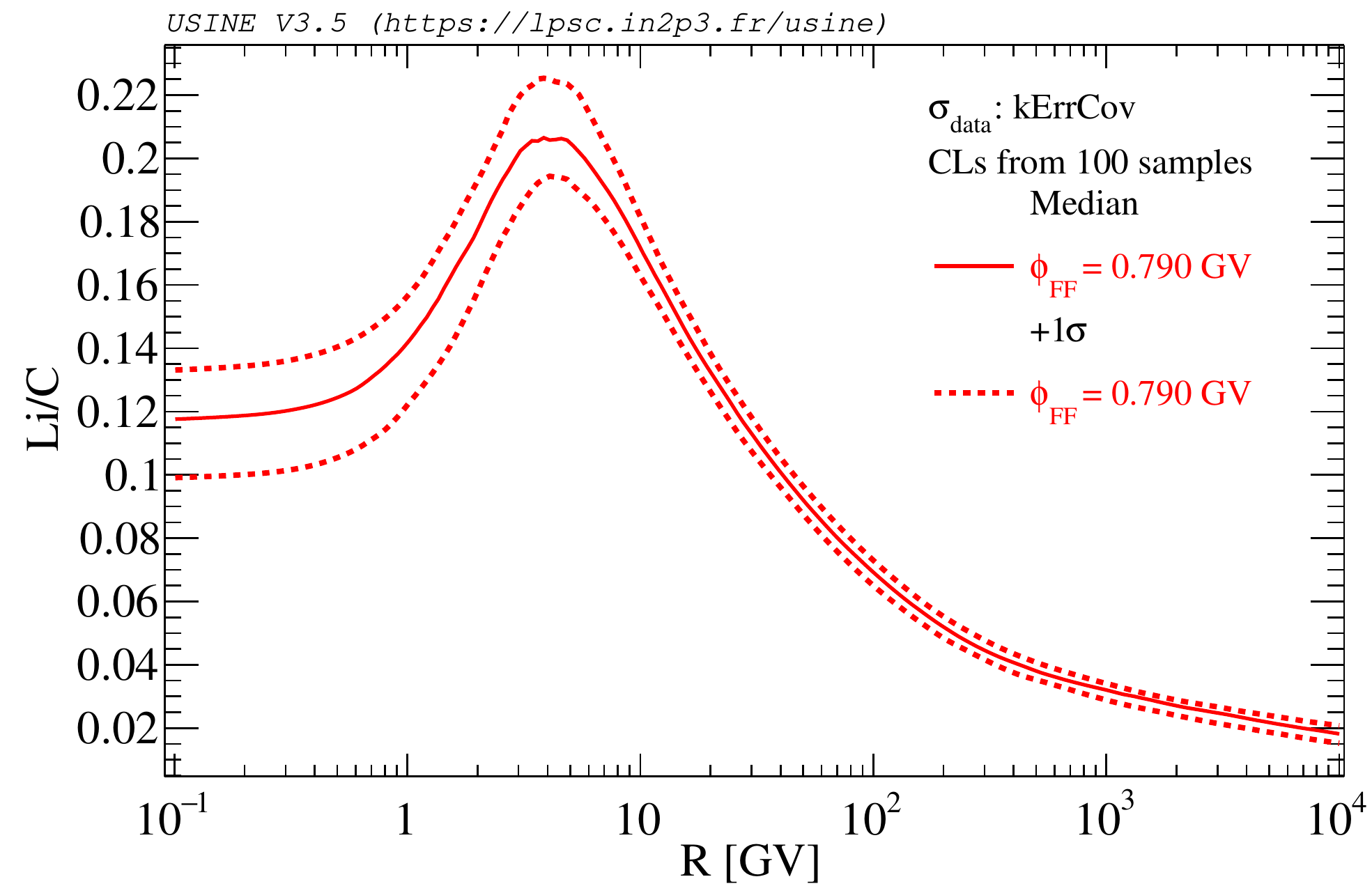}
\caption{Plots from option {\tt -u}. The top panel shows the probability density function of two parameters from the best-fit and covariance matrix of best-fit parameters of model {\sc big} with AMS-02 B/C data \cite{2019arXiv190408917G}, here for 10000 drawn values (option {\tt -u1}). Bottom panel shows $1\sigma$ contours calculated from 100 drawn values of the same transport parameters for Li/C (option {\tt -u2}).}
\label{fig:option_u}
\end{figure}

\begin{itemize}
  \item Nuclear inelastic cross sections ({\tt inputs/XS\_NUCLEI/}) on He for the files {\tt sigInelBarashenkov94.dat}, {\tt sigInelWellish97.dat}, and {\tt sigInelWellish97.dat}  are now correctly based on \cite{1988PhRvC..37.1490F}. I also removed {\tt sigInelLetaw83.dat} because it is an older and less accurate version of {\tt sigInelWellish97.dat}.

  \item New initialisation parameters for minimizer strategy and to save in \ascii{} files the covariance matrix of best-fit parameters (and Hessian).

  \item New prefixes to enable linear combination of cross section nuisance parameters (as described in \cite{2019arXiv190408210D}) and further flexibility in the selection of reactions.

  \item New keyword FIXED and unlimited range enabled for FIT and NUISANCE if minimiser allows.

  \item Source parameters (slope, normalisation, etc.) can now be SHARED (same for all CRs), PERCR (one per CR), PERZ (one per element), or LIST (any combination).

  \item Residuals and score plotted for minimisation, and plots of scans, profile likelihood, and contours enabled (with {\tt minuit}) for FIT and NUISANCE parameters.

  \item For {\tt ./bin/usine -e} runs, the detailed values of all the model free parameters are now shown in a display, and outputs results are now saved in \ascii{} files.

  \item New {\tt ./bin/usine -u} to (i) show 1D and 2D probability density functions drawn from the covariance matrix of best-fit parameters, and (ii) show median and confidence levels on any quantity, as calculated from drawing samples of parameters from the best-fit values and covariance matrix, see Fig.~\ref{fig:option_u}. This option is only possible if the best-fit and covariance matrix of best-fit parameters were obtained previously with the {\tt -m2} option).

  \item Refactoring of several function in {\tt TURunPropagation} to (i) separate calculation step to display step (new class {\tt TURunOutputs} is used to store results and plot them); (ii) break-down $\chi^2$ function into more readable functions ($\chi^2$ from nuisance, $\chi^2$ from data, etc.).

  \item New {\tt ./bin/usine\_pbar} executable to propagate transport, source, and cross-section uncertainties to antiprotons. This was developed and used for the analysis in \cite{2019arXiv190607119B}.

\end{itemize}

\bibliography{usine_v3.5}
\end{document}